\documentclass[]{spie}  

\usepackage{amsmath,amsfonts,amssymb}
\usepackage{graphicx}
\usepackage[most]{tcolorbox}
\usepackage{xcolor}
\usepackage[colorlinks=true, allcolors=blue]{hyperref}

\title {Emergent Physical Intelligence in Biomimetic Scale Metabeams}
\author[a]{Omid Bateniparvar}
\author[a*]{Ranajay Ghosh}
\affil[a]{\textit{Department of Mechanical and Aerospace Engineering, University of Central Florida, Orlando, FL, USA}}

\authorinfo{Further author information: (Send correspondence to Ranajay Ghosh)\\ *E-mail: ranajay.ghosh@ucf.edu}

\begin{document} 
\maketitle

\begin{abstract}
Physical reservoir computing (PRC) leverages the intrinsic dynamics of physical systems to perform information processing while requiring training only in a linear readout layer. Here, we introduce a geometry-programmable metabeam inspired by the hierarchical overlapping architecture of biological scales as a physical reservoir. Contact-mediated interactions between embedded scales induce tunable nonlinear dynamics, enabling controlled transitions between periodic, multi-periodic, and chaotic responses under external excitation. The computational capability of the metabeam reservoir is evaluated using static nonlinear function approximation, Lorenz-63 prediction, and the NARMA-2, NARMA-5, and NARMA-10 benchmarks. Input signals are encoded through the excitation amplitude, whereas computation is realized through the vibrational amplitude of the metabeam followed by a trained linear readout. Across all benchmark tasks, the proposed metabeam consistently outperforms an equivalent linear beam reservoir, yielding lower normalized root-mean-square errors (NRMSEs) and improved computational performance. Different dynamical regimes of the metabeam exhibit distinct computational advantages, with the periodic regime providing the highest prediction accuracy for memory-intensive tasks and the multi-periodic regime achieving the best performance in static nonlinear function approximation. These findings demonstrate that simple interacting surface textures can simultaneously enrich reservoir dynamics and information processing capability, establishing scale-covered metabeams as tunable and mechanically programmable platforms for embodied physical intelligence and PRC.
\end{abstract}


\section{INTRODUCTION}
\label{sec:intro} 
Physical reservoir computing (PRC) is a subfield of artificial intelligence for function approximation and prediction that uses the natural dynamics of physical systems to process information \cite{nakajima2020physical,tanaka2019recent,stepney2024physical}. Instead of solving a problem entirely through programmed digital calculations confined to a dedicated processing unit, PRC uses the physical system itself as the computational medium. Because physical systems evolve continuously over time, their responses depend not only on the current input but also on recent past inputs, providing an inherent fading memory that has been shown to enable temporal information processing in neural networks such as liquid state machines or echo state networks \cite{soures2019deep,jaeger2001short}. These evolving responses, referred to as the reservoir states, are subsequently used to estimate the desired output through a trained readout layer. A wide range of physical systems can act as reservoirs, including optical devices \cite{duport2012all, denis2018all}, electronic circuits \cite{liang2024physical, nowshin2020recent}, fluidic systems \cite{fernando2003pattern,heuthe2026reservoir}, and mechanical structures \cite{shirmohammadli2024physics,coulombe2017computing,dion2018reservoir,tanaka2019recent}. Unlike conventional neural networks (NNs), which typically require optimizing a large number of trainable weights, PRC keeps the physical system unchanged and learns only a simple relationship between the observed responses and the desired output. In practice, input signals are applied to the physical reservoir, the resulting responses are measured over time, and a simple mathematical model converts these responses into the final output. By relying on the natural physics of the system to perform most of the information processing, PRC reduces the amount of training required while maintaining the ability to capture complex nonlinear relationships between inputs and outputs.

Not all physical systems function equally well as reservoirs. For a physical system to perform computation effectively, it must satisfy several key requirements. First, the system must respond differently to different inputs so that useful information can be extracted from its behavior. In reservoir computing (RC), this capability, commonly referred to as the separation property, is achieved by mapping input signals into high-dimensional reservoir states, where nonlinear dynamics generate rich internal representations that make different inputs easier to distinguish \cite{maass2002real}. Second, the system should retain information about recent inputs while gradually reducing the influence of older ones. This property, commonly referred to as fading memory \cite{jaeger2004harnessing,jaeger2001short}, allows the reservoir to process temporal information without indefinitely preserving past history and is often quantified using measures such as memory capacity (MC) \cite{dambre2012information}. Closely related to this requirement is the echo state property (ESP) \cite{jaeger2001echo, gallicchio2017echo}, which ensures that the long-term behavior of the reservoir is determined primarily by the input signal rather than by the initial state of the system. Consequently, two identical reservoirs starting from different initial conditions but receiving the same input will eventually produce similar responses. The balance between high-dimensional state representation and fading memory ultimately governs reservoir performance \cite{tanaka2019recent}, since insufficient nonlinearity limits richness of the reservoir states, whereas excessively complex dynamics may compromise memory retention and reduce predictive accuracy.

Among the various physical platforms explored for RC, mechanical systems have emerged as an attractive platform because they can integrate sensing, actuation, and information processing within a single physical structure \cite{nakajima2020physical,tanaka2019recent,austin2026physical,barazani2020microfabricated}. Unlike digital implementations that separate computation from physical embodiment, mechanical reservoirs perform computation directly through their intrinsic dynamical responses \cite{hauser2011towards,nakajima2013soft}. This capability makes them particularly promising for low-power and embodied intelligence applications, including soft robotics, adaptive structures, and autonomous sensing systems \cite{terajima2025multifunctional,caluwaerts2013locomotion,furuta2018macromagnetic}. Existing mechanical reservoir implementations have demonstrated computational functionality using a variety of physical mechanisms, including material nonlinearity, structural deformation, inertial effects, and compliant structural networks \cite{nakajima2015information,dion2018reservoir,urbain2017morphological}. Early studies primarily exploited the intrinsic nonlinear dynamics of vibrating mechanical structures to generate the rich state representations required for reservoir computing. Representative examples include micromelectroechanical system (MEMS)-based nonlinear resonators \cite{coulombe2017computing,sun2021novel}, sensor-reservoir devices \cite{barazani2020microfabricated}, and networks of coupled mechanical oscillators \cite{sun2021novel,shougat2024multiplex}. More recently, mechanical reservoir computing has expanded toward architected structures such as origami reservoirs, where geometric folding patterns provide programmable nonlinear dynamics for robotic control, perception, and multifunctional information processing \cite{bhovad2021physical,wang2023building}, as well as geometrically nonlinear metamaterials in which computational behavior is governed by structural geometry, topology and internal interactions rather than by material nonlinearity alone \cite{kiyabu2026geometric,he2026embodying,zhang2022harnessing}. These developments indicate a transition from primarily exploiting passive mechanical dynamics to actively engineering computational functionality through structural design. However, in many existing implementations, the reservoir characteristics remain largely governed by fixed material properties, limiting the ability to systematically tune the balance between nonlinear dynamics and fading memory. Unlike intrinsic material nonlinearities, nonlinearities from contact interactions can be systematically programmed through structural geometry while preserving the underlying material properties. Geometry-driven approaches therefore provide an alternative route for engineering reservoir dynamics, allowing computational behavior to emerge directly from structural design and internal interactions. Such geometric programmability creates opportunities to systematically tune reservoir characteristics, expand the accessible dynamical regimes, and establish explicit structure–computation relationships without modifying the underlying material system \cite{kiyabu2026geometric, he2026embodying}.


\begin{figure}[t]
\begin{center}
\begin{tabular}{c} 
\includegraphics[width=1.0\linewidth]{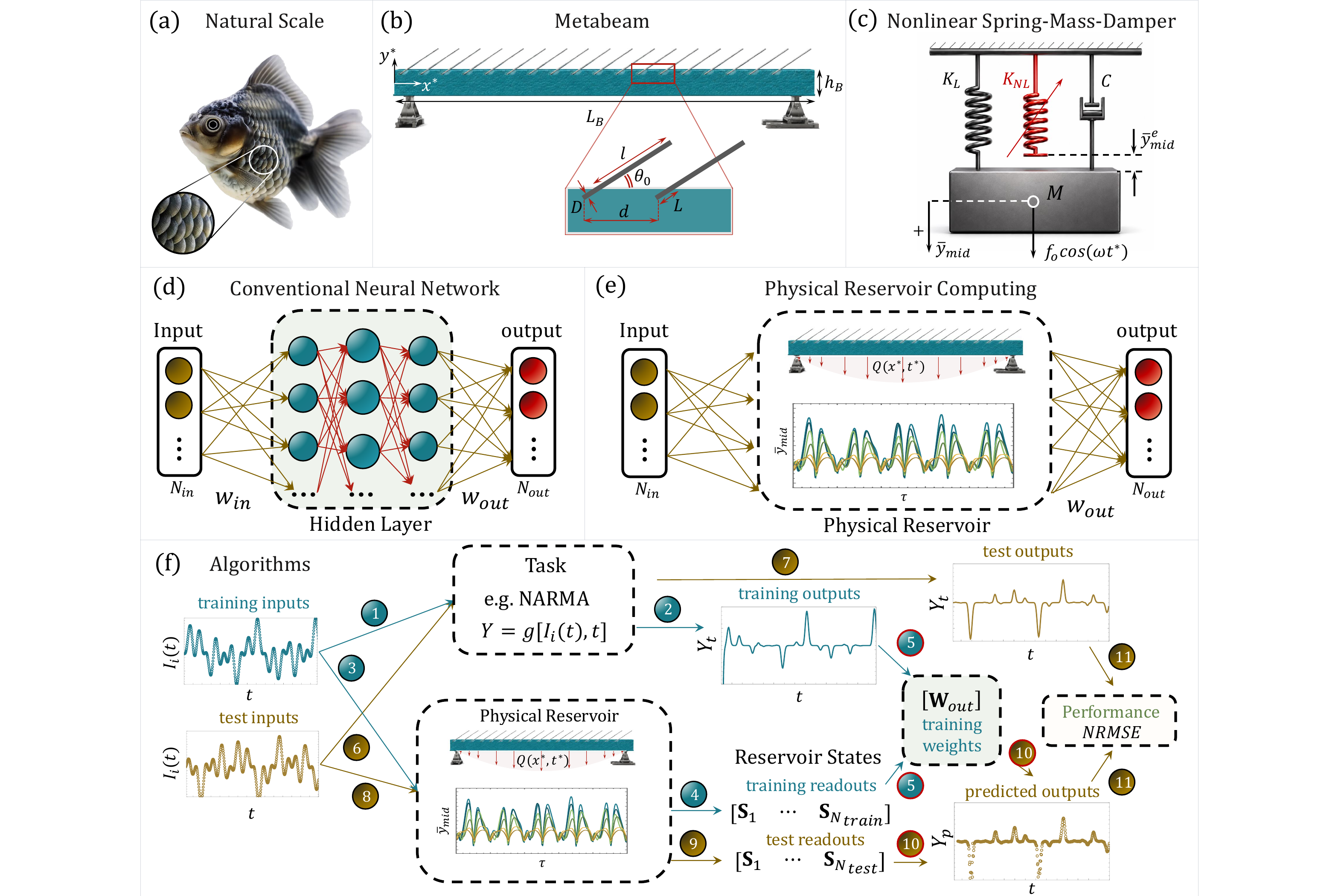}
\end{tabular}
\end{center}

\caption 
{\label{Fig_1}
{\textbf{Metabeam as a physical reservoir for nonlinear computation.}
(a) Biological inspiration from the scale architecture of the goldfish (\textit{Carassius auratus}) \cite{pixabay_fish}, highlighting the overlapping surface structures. (b) Schematic of the metabeam and its geometrical parameters. (c) Reduced-order nonlinear spring–mass–damper (nSMD) representation of the metabeam dynamics, incorporating geometric nonlinearity, contact-mediated stiffness, and dissipation. (d) Conventional neural network (NN) architecture, in which both internal and output weights are optimized through backpropagation. (e) Physical reservoir computing (PRC) paradigm, where the metabeam itself serves as a nonlinear dynamical structure that transforms low-dimensional inputs into rich displacement responses, while only the readout weights are trained. (f) Computational workflow of the proposed framework. Time-varying input signals are encoded into the metabeam, generating high-dimensional reservoir states that are linearly combined through trained readout weights to predict target outputs (e.g., Lorenz and NARMA tasks). Prediction accuracy is quantified using the normalized root-mean-square error (NRMSE).}}

\end{figure} 


Biological surface architectures provide a broad class of mechanically adaptive systems capable of generating complex responses through geometry and local interactions \cite{fratzl2007nature}. Examples include overlapping scales \cite{meyers2008biological}, spines \cite{gorb2008biological}, tongue papillae \cite{noel2018tongue}, feathers \cite{chuong2000evo}, and fur \cite{krsmanovic2023flutter}, all of which exploit structural organization to regulate stiffness, contact, deformation, and environmental interaction. Motivated by this broader class of geometry-mediated mechanisms \cite{vincent2006biomimetics}, we focus here on fish-scale-inspired architectures (Fig.~\ref{Fig_1}(a)) \cite{ghosh2014contact, hossain2022fish} and propose a geometry-programmable metabeam as a physical reservoir for embodied computation. The proposed system consists of an elastic beam integrated with distributed stiff (nearly rigid) plate-like scales that interact through contact during deformation, generating tunable nonlinear dynamics without altering the underlying material properties. Previous work demonstrated that such scale-covered structures can access rich dynamical behaviors, including transitions across periodic, multi-periodic, and chaotic regimes \cite{bateniparvar2024dynamic,bateniparvar2026chaotic, bateniparvar2026biomimetic}. Input signals are encoded into the metabeam through excitation, and the resulting vibrational responses are sampled over time to construct reservoir states. These states are subsequently combined through a linear readout layer to estimate the desired output. By controlling geometric parameters and excitation conditions, the metabeam enables systematic adjustment of the balance between nonlinear dynamics and fading memory. To evaluate its computational capability, the proposed framework is examined using nonlinear function approximation, prediction of the Lorenz-63 chaotic system \cite{lorenz2017deterministic,gauthier2021next}, and Nonlinear Autoregressive Moving Average (NARMA) prediction tasks \cite{atiya2000new,he2025role} while comparing its performance against an equivalent linear elastic beam reservoir. Using this framework, we establish how geometry-induced contact interactions influence computational performance and memory characteristics in PRC.

Taken together, the proposed framework establishes a route toward geometry-programmable PRC, where computational behavior emerges from contact interactions rather than from modifying material composition or increasing algorithmic complexity. Beyond demonstrating the feasibility of metabeams as physical reservoirs, this study reveals that computational performance is strongly influenced by the underlying dynamical regime and that greater dynamical complexity does not necessarily imply superior computational capability. Instead, different dynamical regimes provide distinct computational advantages depending on the memory and nonlinear processing demands of the target task. These findings demonstrate that geometry as an active design parameter for programming physical computation and suggest new opportunities for developing mechanically embodied intelligent systems that integrate structure, dynamics, and information processing within a unified physical platform.

\section{Methods and Models}

\subsection{Physical Reservoir System}

This section presents the proposed contact-mediated metabeam reservoir, its singular reduced-order dynamical model (sROM), input encoding strategy, and the reservoir state-generation framework used throughout the subsequent benchmark tasks.

\subsubsection{Metabeam as Nonlinear Dynamical System}

The proposed metabeam consists of a slender elastic substrate integrated with a periodic array of rigid scales, as illustrated in Fig.~\ref{Fig_1}(b). The substrate is defined by its length $L_B$ and height $h_B$, whereas each scale is defined by its length $l$, thickness $D$, spacing $d$, and initial inclination angle $\theta_0$. The overlap ratio, $\eta=l/d$, quantifies the geometric overlap between neighboring scales and serves as a primary geometric parameter governing contact engagement and the resulting nonlinear dynamics.

As the substrate bends, the attached scales rotate with the beam. Prior to the onset of contact engagement, the metabeam behaves as a linear beam. As the deformation exceeds the geometric engagement threshold, neighboring scales progressively come into unilateral contact, introducing an additional restoring mechanism that modifies the effective stiffness. Because this process is governed by geometry-mediated contact rather than intrinsic material nonlinearity, the resulting dynamical behavior can be systematically programmed through geometric design. The kinematic and dynamic foundations of scale-covered metabeam interaction have been established in prior studies \cite{ghosh2014contact,sarkar2025bending,ali2019frictional,ebrahimi2023material,bateniparvar2024dynamic} and therefore are not reproduced here in full detail. Instead, the present work adopts a reduced-order dynamical formulation developed in our recent work \cite{bateniparvar2026chaotic}. The metabeam is modeled as a simply supported flexural system and reduced using a first-mode approximation, resulting in an equivalent nonlinear spring–mass–damper (nSMD) model shown in Fig.~\ref{Fig_1}(c).

The governing equations are expressed in nondimensional form by scaling the horizontal coordinate $x^*$ with the beam length $L_B$, the transverse displacement $y^*$ with the beam height $h_B$, and time with the first natural frequency $\omega_0$ of the corresponding linear simply supported beam. Accordingly, $\bar{x}=x^*/L_B$, $\bar{y}=y^*/h_B$, $\tau=\omega_0 t^*$, and $\Omega=\omega/\omega_0$, where $\omega_0=\pi^2\sqrt{EI/(\rho S L_B^4)}$. Here, $E$ is the Young's modulus of the substrate, $I$ is the cross-sectional moment of area, $\rho$ is the substrate mass density, and $S$ is the cross-sectional area.

The resulting sROM describes the nonlinear dynamics of the metabeam under harmonic excitation. The governing nondimensional equation of motion is given by \cite{bateniparvar2026chaotic,ali2019frictional}

\begin{equation}
\label{eq1}
\ddot{\bar{y}}_{mid}(\tau)+\delta\dot{\bar{y}}_{mid}(\tau)+\bar{y}_{mid}(\tau)+\overline{F}_{NL}\left(\bar{y}_{mid};\bar{K},\eta,\theta_0,\zeta,\bar{d}\right)
=\gamma\cos(\Omega\tau),
\end{equation}

where $\bar{y}_{mid}$ denotes the normalized transverse displacement at metabeam midspan, $\delta$ is the damping ratio, $\Omega$ is the excitation frequency normalized by the first natural frequency of the beam $\omega_0$, and $\gamma$ is the normalized forcing amplitude used to encode the input signal into the reservoir. The nonlinear restoring term $\overline{F}_{NL}$ represents the contact-mediated restoring force arising from progressive scale engagement and is governed by the geometric design parameters, including the normalized torsional stiffness $\bar K$, overlap ratio $\eta$, initial inclination angle $\theta_0$, nondimensional beam slenderness $\zeta=h_B/L_B$, and normalized scale spacing $\bar d=d/L_B$. The solution of Eq.~\ref{eq1} over each excitation interval provides the continuous dynamical response, from which the reservoir states are obtained through temporal sampling.

\subsubsection{Reservoir State Generation}

To generate a reservoir state for each encoded input (e.g., a segment of a vibration signal, a sensor measurement, or any other time-varying signal), the metabeam response is computed over a finite excitation interval. For each excitation interval, the metabeam evolves over a finite temporal horizon of duration $T_c$, defined as

\begin{equation}
\label{eq2}
\bar T=\frac{2\pi}{\Omega},\qquad 
T_c=n_c\bar T,
\end{equation}

where $n_c$ denotes the number of forcing cycles. The continuous displacement response over the interval $[0, T_c]$ is then sampled at $R$ prescribed observation instants $\tau_r\in[0,T_c]$, where $r=1,\ldots,R$. Sampling the response at multiple observation instants captures the temporal evolution of the metabeam within each excitation interval and forms the reservoir-state vector:


\begin{equation}
\label{eq4}
\mathbf S_i
=
\left[
\bar{y}_{mid}(\tau_1),
\bar{y}_{mid}(\tau_2),
\ldots,
\bar{y}_{mid}(\tau_R)
\right]^T
\in
\mathbb R^{R\times 1},
\end{equation}

where $\mathbf S_i$ denotes the sampled state of the metabeam dynamics (in this case, midpoint deflection) associated with the $i^{th}$ excitation interval. Each excitation interval therefore produces one reservoir-state vector, and the collection of all state vectors forms the reservoir-state matrix

\begin{equation}
\label{eq5}
\mathbf {U}=\left[\mathbf {S}_1,\mathbf {S}_2,\ldots,\mathbf {S}_N\right]\in\mathbb R^{R\times N},
\end{equation}

which defines the observable reservoir state space used for downstream learning, where $N$ denotes the total number of reservoir states (equivalently, the total number of encoded input intervals). Each column of $\mathbf U$ therefore represents the reservoir response to one encoded input interval. Unlike conventional digital reservoirs, where memory arises through recurrent algorithmic state evolution, memory in the proposed metabeam emerges intrinsically from the continuous system dynamics. Inertial effects, damping-controlled fading dynamics, and nonlinear contact interactions collectively preserve information across successive excitation intervals, transforming temporally encoded inputs into a high-dimensional dynamical representation. These reservoir states constitute the feature representation used by the output readout in the subsequent learning stage.

\subsubsection{Input Encoding and Reservoir Excitation}

Having established the governing dynamics of the proposed metabeam reservoir, the next step is to define how external information is introduced into the system. To enable temporal information processing, external inputs are encoded into controllable reservoir variables that interact with the physical reservoir.

From a general continuum dynamical perspective, information may in principle be injected into the reservoir through any parameter participating in the governing equation, $\mathcal I\rightarrow\Theta$, where $\mathcal I$ denotes the input space and $\Theta$ represents the set of controllable reservoir parameters. For the present metabeam, candidate encoding channels include geometric parameters (e.g., overlap ratio $\eta$, and initial scale inclination $\theta_0$), dissipative parameter (e.g., damping ratio $\delta$), and excitation parameters associated with external forcing. 

Although geometric and dissipative parameters are theoretically admissible encoding variables, modifying them during operation requires either physical reconfiguration or changes to the intrinsic system dynamics, making them unsuitable for continuous temporal computation. Accordingly, the present work adopts forcing-based encoding, in which information is injected exclusively through modulation of excitation amplitude $\gamma$ while preserving the intrinsic mechanical characteristics of the reservoir. The excitation process is therefore represented as a sequence of piecewise-constant forcing amplitudes

\begin{equation}
\gamma(\tau)=\gamma_i,\qquad\forall\;\tau\in[(i-1)T_c,iT_c],
\end{equation}

where $\gamma_i$ denotes the encoded forcing amplitude associated with the $i^{th}$ excitation interval. The encoded forcing amplitude is obtained through the mapping $\gamma_i=G(I_i)$, where $I_i\in\mathbb R$ denotes the input state associated with the $i^{th}$ excitation interval and $G(\cdot)$ represents the encoding operator. Throughout this work,  the excitation frequency is held fixed ($\Omega=\Omega_0$), such that information is injected exclusively through forcing amplitude modulation while preserving the intrinsic reservoir dynamics and maintaining a consistent computational substrate. 
 
 In the present implementation, the encoding operator is restricted to bounded amplitude modulation to preserve operation within dynamically informative regions of the reservoir response. Accordingly, a linear amplitude encoding strategy is adopted as

\begin{equation}
\gamma_i
=
\gamma_{\min}
+
(\gamma_{\max}-\gamma_{\min})
\frac{
I_i-I_{\min}
}{
I_{\max}-I_{\min}
},
\end{equation}

where $\gamma_{\min}$ and $\gamma_{\max}$ define the admissible forcing interval and $[I_{\min},I_{\max}]$ denotes the input domain. The selected forcing bounds maintain reservoir operation within dynamically informative regimes that produce distinguishable reservoir states. Because the contact-mediated metabeam exhibits periodic, multiperiodic, and increasingly complex oscillatory responses over different forcing amplitudes, restricting the excitation interval avoids state saturation and ensures reliable reservoir operation throughout both training and inference. 

Under the adopted encoding strategy, information is injected solely through forcing amplitude modulation, while all intrinsic reservoir parameters remain fixed throughout operation. The metabeam is evolved continuously over the complete input sequence without state reset. Consequently, both the displacement and velocity remain continuous across successive excitation intervals, $\bar{y}_{mid}(iT_c)^+=\bar{y}_{mid}(iT_c)^-$ and $\dot{\bar{y}}_{mid}(iT_c)^+=\dot {\bar{y}}_{mid}(iT_c)^-$ where superscripts $-$ and $+$ denote the states immediately before and after the excitation interval boundary, respectively. This continuous evolution preserves temporal continuity and allows the intrinsic physical memory of the reservoir to influence subsequent state evolution. The resulting encoding–evolution process may therefore be interpreted as

\begin{equation}
\label{eq8}
I_i\rightarrow\gamma_i\rightarrow \bar{y}_{mid}(\tau)\rightarrow\mathbf S_i\rightarrow\mathbf U,
\end{equation}

which summarizes the complete physical state-generation pipeline used for downstream learning tasks and corresponds to the encoding and reservoir state generation stages.

\subsection{Learning Problem and Readout}

Once the physical reservoir has transformed encoded inputs into high-dimensional state representations, the target output is inferred through supervised learning. Unlike conventional NNs, where learning requires iterative optimization of numerous internal parameters, PRC confines training to a linear output readout while leaving the reservoir dynamics unchanged. This section formulates the learning problem, presents the readout training procedure, and introduces the performance metrics used to evaluate the computational capability of the proposed metabeam reservoir.

\subsubsection{Learning Problem Formulation}

The reservoir states generated in the previous section provide a nonlinear dynamical representation of the encoded input sequence, but do not directly represent the desired computational output. A supervised learning procedure is therefore used to establish a mapping from the observable reservoir states to the target outputs.

The overall objective of the proposed physical reservoir system is to approximate an unknown temporal input--output relationship. This approximation is achieved in two stages. First, the encoded input sequence is transformed by the metabeam dynamics into a sequence of observable reservoir states. Second, a trainable output readout maps these reservoir states to the target output space. The overall input--output mapping may therefore be expressed as $\mathcal I\rightarrow \mathcal Y$ where $\mathcal I$ and $\mathcal Y$ denote the input and target output spaces, respectively. Within this overall mapping, however, the metabeam itself is not trained, only the output readout is estimated from data. This distinction separates the proposed framework from conventional neural architectures, as illustrated in Fig.~\ref{Fig_1}(d), in which both input and output weights are iteratively optimized during training. 

In the proposed physical reservoir framework, Fig.~\ref{Fig_1}(e), the intrinsic metabeam dynamics perform the nonlinear state transformation, while learning is restricted to the estimation of a linear output readout. For each encoded input value $I_i$, the metabeam generates a corresponding reservoir state $\mathbf S_i$ (Eq.~\ref{eq4}). The complete reservoir-state matrix $\mathbf U$ (Eq.~\ref{eq5}) is paired with the target sequence

\begin{equation}
\label{eq9}
\mathbf Y=[Y_1,Y_2,\ldots,Y_N]\in\mathbb R^{1\times N},
\end{equation}

where $Y_i$ denotes the target output associated with the $i^{\mathrm{th}}$ excitation interval. Thus, each column $\mathbf{S}_i$ of $\mathbf{U}$ is paired with one target value $Y_i$. 

Before readout estimation, the available state-target pairs are partitioned into training and testing subsets. During the training stage, the training reservoir matrix $\mathbf U$ and its corresponding target outputs $\mathbf Y$ are used to estimate the output weight $\mathbf W_{out}$. Once estimated, $\mathbf W_{out}$ remains fixed. During the prediction stage, the trained readout is applied to previously unseen testing states to generate the predicted output sequence $\mathbf{Y}_{p}$. This overall workflow is illustrated schematically in Fig.~\ref{Fig_1}(f).

\subsubsection{Readout Estimation}

Within the RC paradigm, learning is performed exclusively through estimation of the output readout, while the internal reservoir dynamics remain fixed. To estimate the output readout, the reservoir-state matrix obtained from the training data is first standardized component-wise using normalization parameters computed exclusively from the training set. This standardization places all reservoir-state components on comparable numerical scales and improves the stability of the subsequent readout estimation. Denoting the corresponding raw reservoir-state matrix by $\mathbf U$, the augmented normalized state matrix is constructed as

\begin{equation}
\label{eq10}
\tilde{\mathbf U}
=
\left[
\frac{\mathbf U-\boldsymbol\mu}{\boldsymbol\sigma},
\quad \mathbf 1
\right]^T
\in
\mathbb R^{(R+1)\times N},
\end{equation}

where $\boldsymbol\mu$, $\boldsymbol\sigma$ $\in
\mathbb R^{R\times1}$ denote the component-wise mean and standard deviation computed from the training states, respectively, and $\mathbf 1\in\mathbb R^{1\times N}$ denotes a bias row appended to account for affine offsets in the output mapping.

The output readout is obtained by solving a linear least-squares optimization problem. Although this optimization problem can be solved using the Moore-Penrose pseudoinverse \cite{lukovsevivcius2012practical,penrose1955generalized}, the resulting solution may become numerically unstable when the reservoir-state matrix is poorly conditioned. Therefore, a ridge-regularized least-squares formulation \cite{lukovsevivcius2012practical,hoerl2000ridge} is adopted throughout this work. The corresponding closed-form solution is

\begin{equation}
\label{eq11}
\mathbf W_{out}=
\mathbf Y
\tilde{\mathbf U}^{T}
\left(
\tilde{\mathbf U}
\tilde{\mathbf U}^{T}
+
\lambda
\mathbf {\tilde{R}}
\right)^{-1}
\in
{\mathbb {R}^{1\times (R+1)}},
\end{equation}

where $\lambda>0$ is the ridge regularization parameter used to improve numerical conditioning and reduce overfitting, and $\mathbf {\tilde{R}}=\mathrm{diag}(1,\ldots,1,0)\in\mathbb R^{(R+1)\times(R+1)}$ is the diagonal regularization matrix, with the bias term excluded from regularization. Throughout this work, a fixed value of $\lambda=10^{-5}$ is used for all reported results. Once the output readout has been estimated, it remains fixed throughout the prediction stage. The testing reservoir-state matrix $\mathbf U^*$ are standardized according to the normalization defined in Eq.~\ref{eq10} and augmented using the identical bias construction, yielding the normalized testing state matrix $\tilde{\mathbf U}^*$. The predicted output sequence is then obtained as

\begin{equation}
\label{eq12}
\mathbf Y_{p}=
\mathbf W_{out}
\tilde{\mathbf U}^*.
\end{equation}

\subsubsection{Performance Evaluation}

Following readout estimation, the predictive performance of the proposed physical reservoir is evaluated using previously unseen testing data. Two complementary performance metrics are employed: the normalized root mean square error (NRMSE), which quantifies prediction accuracy, and the coefficient of determination ($R^2$), which measures goodness of fit. Prediction accuracy is first evaluated using NRMSE, defined as

\begin{equation}
\label{eq13}
\mathrm{NRMSE(\%)}=
\frac{
\sqrt{
\frac{1}{N}
\sum_{i=1}^{N}
\left(
Y_i-\hat Y_i
\right)^2
}
}{
Y_{\max}-Y_{\min}
}\times100,
\end{equation}

where $Y_i$ and $\hat {Y_i}$ denote the target and predicted outputs, respectively, and $Y_{\max}$ and $Y_{\min}$ represent the maximum and minimum values of the target sequence. NRMSE is reported as a percentage relative to the target dynamic range. To further quantify goodness of fit, the coefficient of determination $R^2$ is additionally evaluated as

\begin{equation}
\label{eq14}
R^2=
1-
\frac{
\sum_{i=1}^{N}
(Y_i-\hat Y_i)^2
}{
\sum_{i=1}^{N}
(Y_i-\bar Y)^2
}
\end{equation}

where $\bar Y$ denotes the mean target value and $R^2=1$ corresponds to perfect agreement between the predicted and target outputs. Together, NRMSE and $R^2$ provide complementary measures of prediction accuracy and goodness of fit, corresponding to the final evaluation stage.

\subsubsection{Memory Capacity Evaluation}

Memory capacity (MC) characterizes the ability of the physical reservoir to retain previously encoded input information and make it available for subsequent computation \cite{jaeger2001short,stepney2024physical}. An effective reservoir should preserve sufficient information from recent inputs while gradually forgetting older ones, thereby providing the fading-memory property required for temporal information processing \cite{jaeger2001short,maass2002real}. Insufficient memory limits the influence of relevant past inputs, whereas excessively long memory allows obsolete information to persist, degrading the fading-memory property and may reduce the reservoir's ability to generalize \cite{cucchi2022hands}.

To quantify this property, delayed-input reconstruction was performed over increasing memory depths. The memory depth, denoted by $m$, represents the number of excitation intervals separating the current reservoir state from the encoded input to be reconstructed. For each memory depth $m$, an independent ridge-regression readout was trained to reconstruct the encoded input applied $m$ excitation intervals earlier from the current reservoir state.

The memory contribution at memory depth $m$ was quantified using the squared Pearson correlation coefficient between the reconstructed and reference delayed inputs \cite{stepney2024physical},

\begin{equation}
MC_m=
\frac{
\mathrm{cov}^{2}
\left(
I_{i-m},\hat{I}_{i-m}
\right)
}{
\mathrm{var}
\left(
I_{i-m}
\right)
\,
\mathrm{var}
\left(
\hat{I}_{i-m}
\right)
},
\end{equation}

where $I_{i-m}$ and $\hat{I}_{i-m}$ denote the reference and reconstructed delayed inputs, respectively. The resulting memory contribution satisfies $0 \le MC_m \le 1$, with larger values indicating more accurate reconstruction of information retained over the corresponding memory depth.

The overall memory capacity was computed by summing the memory contributions over all evaluated memory depths \cite{stepney2024physical},

\begin{equation}
MC=\sum_{m=1}^M MC_m,
\end{equation}

where $M$ denotes the maximum evaluated memory depth. The memory profile $MC_m$ describes how the reservoir's ability to reconstruct previously encoded inputs decays with increasing memory depth, whereas the cumulative quantity $MC$ provides a scalar measure of the total recoverable input history. For each memory depth, the readout was trained using the same standardized reservoir states and ridge-regression regularization adopted for all computational tasks.

\subsection{Computational Tasks}

Having established the physical reservoir, the learning framework, and the output readout, the final step is to evaluate the computational capability of the proposed system using representative benchmark tasks. An effective physical reservoir should provide rich nonlinear representations, retain relevant temporal information, and generalize across diverse computational tasks. To assess these complementary computational properties, three benchmark tasks are considered. Static nonlinear function approximation evaluates the intrinsic nonlinear representational capability of the reservoir in the absence of temporal memory. The Lorenz-63 benchmark examines nonlinear state reconstruction from continuously evolving chaotic dynamics, while the NARMA benchmarks evaluate the combined ability of the reservoir to perform nonlinear processing together with temporal information retention over progressively increasing memory horizons \cite{wringe2025reservoir,atiya2000new,lorenz2017deterministic}.

\subsubsection{Task Formulation}

To ensure a consistent evaluation framework, all benchmark tasks considered in this study employ the same physical reservoir, input encoding protocol, and learning procedure. Differences between tasks arise solely from the construction of the input and target sequence, allowing the influence of the computational objective to be isolated from changes in the underlying reservoir dynamics. This common formulation provides a unified basis for comparing reservoir performance across different computational settings.

Given an input sequence $\mathbf I=[I_1,I_2,\ldots,I_N]\in
\mathbb R^{1\times N}$, the encoded excitation protocol described previously is applied sequentially to generate the corresponding reservoir-state matrix $\mathbf U$. Task-specific target outputs are subsequently constructed according to the definition of each benchmark. 

For each input $I_i$, the metabeam evolves dynamically over the prescribed excitation horizon and generates the corresponding reservoir state $\mathbf S_i$. Because successive reservoir states remain physically connected through state continuity, each state naturally retains information from previous excitation intervals, thereby encoding temporal dependencies beyond the current input. The computational objective is then formulated as estimating an output readout that maps the reservoir states to the corresponding task-specific target responses. Consequently, temporal processing is not achieved through explicit delay embedding or autoregressive reconstruction, but instead emerges naturally from the continuous evolution of the physical reservoir. All reservoir parameters remain unchanged across the benchmark tasks, and only the construction of the encoded input and target output is modified. For temporal tasks, training and testing reservoir states are generated sequentially while preserving state continuity.

\subsubsection{Non-Temporal Nonlinear Function Approximation}

The first benchmark considers static nonlinear function approximation to evaluate the intrinsic nonlinear representational capability of the proposed metabeam reservoir independently of temporal memory requirements. Unlike the temporal benchmarks introduced subsequently, this task isolates the reservoir's ability to construct nonlinear mappings without requiring reconstruction of delayed dependencies. For each input $I_i$, the encoded excitation protocol described previously is applied to generate the corresponding reservoir state $\mathbf S_i$, and the output readout is trained to approximate a prescribed static nonlinear mapping. Two nonlinear test functions are considered to assess complementary aspects of nonlinear approximation. The first is a high-order polynomial function ($f_1$) adopted from the nonlinear approximation literature \cite{mandal2022machine,arun2024reservoir}, whereas the second is a localized oscillatory nonlinear function ($f_2$) introduced in the present work to complement the evaluation by testing the reservoir's ability to approximate localized nonlinear variations:

\begin{equation}
\label{eq15}
f_1(x)=
(x-3)(x-2)(x-1)x(x+1)(x+2)(x+3),
\qquad
f_2(x)=
10e^{-5x^2}\cos(8x).
\end{equation}


For both nonlinear functions above, prediction performance is evaluated across distinct reservoir operating regimes identified from the dynamical characterization introduced previously, including periodic (P), multi-periodic (MP), and chaotic (C) responses. Prediction accuracy is quantified using the NRMSE metric defined earlier and examined as a function of the training set size $N_{\mathrm{train}}$. These tasks provide a controlled assessment of nonlinear approximation capability while eliminating temporal memory requirements.

\subsubsection{Lorenz-63 Benchmark}

Unlike the static function approximation tasks, the Lorenz-63 benchmark evaluates the proposed reservoir under continuously evolving chaotic dynamics. The objective is to reconstruct nonlinear relationships between observable and hidden system states while exploiting the reservoir's inherent temporal processing capability. The Lorenz-63 system represents a physically meaningful nonlinear dynamical process and therefore provides a complementary assessment of nonlinear state reconstruction in continuous chaotic systems. The Lorenz-63 dynamics are governed by

\begin{equation}
\label{eq16}
\dot x=
\sigma
(y-x),
\qquad
\dot y=
x(\rho-z)-y,
\qquad
\dot z=
xy-\beta z,
\end{equation}

where ($x$, $y$, $z$) denote the state variables and $\sigma$, $\rho$, and $\beta$ are the governing system parameters. In the present study, the canonical parameter set $\sigma=10$, $\rho=28$, $\beta=8/3$ is adopted together with the initial condition $x(0)=1$, $y(0)=1$, $z(0)=1$. The Lorenz system is numerically integrated to generate the reference trajectories. Unlike the NARMA benchmark, where target outputs are generated recursively from an externally prescribed input sequence, the Lorenz benchmark constructs both input and target directly from an underlying continuous dynamical system. Specifically, the observable Lorenz coordinate $x(t)$ is used to define the input states, while the corresponding $z(t)$ trajectory serves as the target output.

Accordingly, the input and target states are defined as

\begin{equation}
\label{eq17}
I_i=x(t_i),
\qquad
Y_i=z(t_i),
\end{equation}

where $t_i$ denotes the discrete sampling instants extracted from the continuous Lorenz trajectory. The resulting input sequence is subsequently encoded into excitation amplitudes according to the reservoir encoding protocol introduced previously. Successful prediction therefore requires the metabeam reservoir to retain temporal information from the observed trajectory while reconstructing nonlinear state relationships associated with the underlying Lorenz dynamics. Although the Lorenz benchmark evaluates temporal processing in a physically generated chaotic system, it does not systematically vary the memory requirement. This aspect is addressed in the following NARMA benchmarks.

\subsubsection{NARMA Benchmark}

To systematically evaluate nonlinear processing together with temporal information retention, NARMA benchmark tasks are considered. NARMA systems are widely adopted in RC because successful prediction requires simultaneous reconstruction of nonlinear interactions and delayed dependencies over multiple temporal horizons. Increasing the NARMA order progressively strengthens the temporal memory requirement while preserving the nonlinear nature of the benchmark, thereby providing a systematic assessment of reservoir performance across different memory depths.

Unlike conventional NARMA implementations that employ independently sampled random inputs, the present study adopts a bounded multiharmonic excitation protocol to maintain controlled operation of the physical reservoir within dynamically informative regimes. The input sequence is generated according to

\begin{equation}
\label{eq18}
I_i=
A
\prod_{j=1}^{3}
\sin
\left(
\alpha_j
\Omega_{exc}
t_i
\right),
\end{equation}

where $\prod$ denotes multiplication of the prescribed harmonic components, $A$ denotes the input amplitude, $\Omega_{exc}$ is the excitation frequency, and $\alpha_j$ defines the selected frequency multipliers. In the present implementation, $A=0.2$, $\boldsymbol{\alpha}=\{0.56,1.00,1.16\}$ and $\Omega_{exc}=1.25$. The generated input sequence is subsequently encoded into the forcing amplitude of the metabeam according to the encoding strategy introduced previously.

For an $n^{\mathrm{th}}$-order NARMA system, the target output is generated as

\begin{equation}
\label{eq19}
Y_i=
\begin{cases}
0.4\,Y_{i-1}
+0.4\,Y_{i-1}Y_{i-2}
+0.6\,I_i^{3}
+0.1,
& n=2 \\[2mm]
0.3\,Y_{i-1}
+0.05\,Y_{i-1}
\displaystyle\sum_{j=i-n}^{i-1} Y_j
+1.5\,I_{i-n}I_{i-1}
+0.1,
& n\in\{5,10\}
\end{cases},
\end{equation}

where $n$ denotes the order of the NARMA system and determines the effective memory requirement of the task. In this study, NARMA orders $n\in\{2,5,10\}$ are considered to systematically evaluate the computational capability of the proposed metabeam reservoir. Lower-order tasks primarily assess high-dimensional nonlinear state representations, whereas increasing temporal order imposes progressively stronger requirements on temporal information retention and delayed nonlinear dependencies. 

Together, these benchmark tasks provide a systematic evaluation of reservoir performance across progressively increasing temporal memory requirements.

\subsection{Numerical Implementation}

Having introduced the physical reservoir, learning framework, and computational benchmarks, the final step is to specify the numerical implementation adopted throughout this study. To ensure reproducibility and facilitate consistent comparison across benchmark tasks, a common reservoir configuration, operating regime, and training protocol are employed unless otherwise stated. This section describes the numerical implementation adopted throughout the study, including the reservoir configuration, operating regime, dataset construction, and training protocol used in all subsequent computational experiments.


\begin{figure}[t]
\begin{center}
\begin{tabular}{c} 
\includegraphics[width=1.0\linewidth]{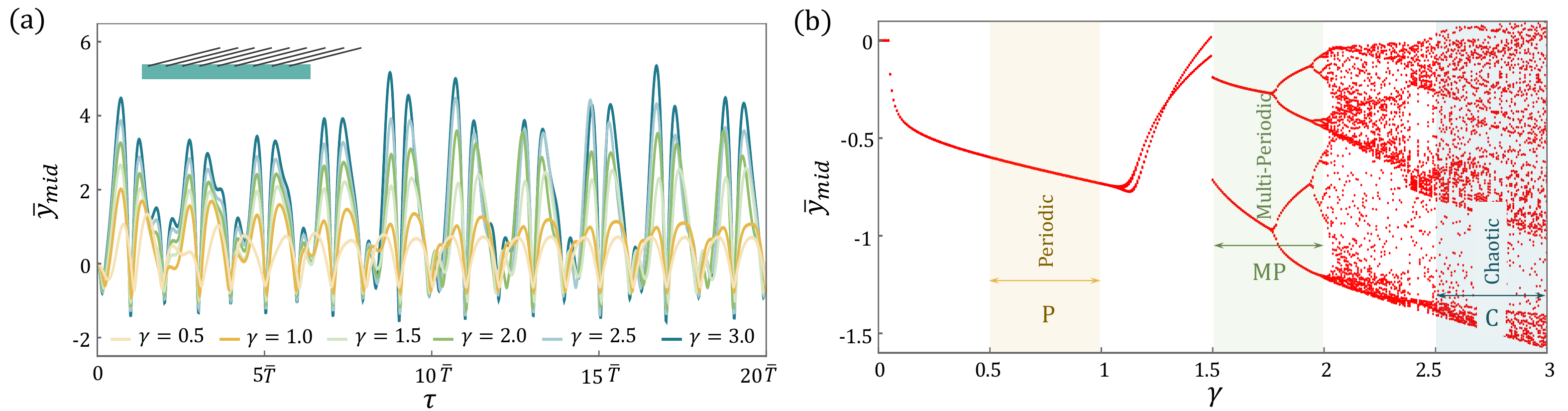}
\end{tabular}
\end{center}

\caption 
{\label{Fig_2}
{\textbf{Time histories and bifurcation characteristics of the metabeam reservoir.} (a) Representative midspan responses for increasing excitation amplitudes $\gamma$. (b) Bifurcation diagram showing the evolution from periodic (P) to multi-periodic (MP) and chaotic (C) dynamics. 
}}

\end{figure} 


\subsubsection{Reservoir Configuration and Operating Regime}

The computational performance of a physical reservoir depends strongly on the operating regime in which its dynamics evolve. To isolate the influence of the computational tasks from changes in the reservoir itself, all computational benchmarks were performed using a fixed metabeam configuration.

The nondimensional reservoir parameters were selected as $\eta=10$, $\theta_0=3^\circ$, $\bar K=4.5$, $\delta=0.1$, $\Omega=1$, $\zeta=0.02$, $\bar d=0.06$ and remained unchanged across all benchmark tasks, including the Lorenz–63 and NARMA benchmarks. These parameters correspond to the validated physical metabeam configuration established in our previous study \cite{bateniparvar2026chaotic}. This parameter set was chosen to position the reservoir within a dynamically informative operating region exhibiting rich nonlinear behavior while maintaining stable state evolution. To characterize the accessible dynamical responses of the metabeam under force-driven excitation, the system behavior was examined over varying excitation amplitudes.

Fig.~\ref{Fig_2} summarizes the dynamical operating regimes of the metabeam under varying forcing amplitudes. The bifurcation analysis was used solely to identify the excitation range adopted for input encoding in the computational benchmarks. Detailed discussion of the underlying nonlinear dynamics is deferred to the Results section.

\subsubsection{Dataset Construction and Training Protocol}

Beyond the fixed reservoir configuration, a consistent numerical protocol is required to ensure fair comparison across all computational benchmarks. For temporal benchmark tasks, reservoir states were generated sequentially by repeatedly evolving the metabeam under the encoded excitation protocol while preserving state continuity between successive excitation intervals. In contrast, for non-temporal function approximation, the reservoir was reinitialized to the prescribed initial condition before each input sample to eliminate temporal effects and isolate the intrinsic nonlinear representation capability of the reservoir.

In this study, each input excitation was applied for 20 forcing cycles ($n_c=20$), with $10$ uniformly distributed sampling instants per cycle, resulting in $R=200$ reservoir observations for each input interval. Sampling was performed over the interval $[0,T_c)$, excluding the terminal endpoint to ensure that all sampled reservoir states corresponded exclusively to the current input excitation. To reduce dependence on the initial reservoir conditions and eliminate transient responses before the reservoir reaches its characteristic response, an initial washout stage was introduced prior to training and evaluation. The first $N_{\mathrm{wash}}=50$ reservoir states were discarded to ensure that the subsequent readout training was based solely on the intrinsic reservoir dynamics.

Because benchmark tasks impose different computational requirements, training and testing datasets were selected individually for each task. Specifically, $\{N_{train},N_{test}\}=\{100,100\}$ for non-temporal function approximation and $\{N_{train},N_{test}\}=\{500,500\}$ for temporal benchmarks (Lorenz–63 and NARMA). The reduced-order metabeam dynamics were solved numerically using $\texttt{NDSolve}$ in $\texttt{Wolfram Mathematica}$ with adaptive time integration. Output readout estimation was performed using linear ridge regression with a regularization parameter of $\lambda=10^{-5}$. Unless otherwise stated, identical numerical settings were maintained across all benchmark tasks to ensure that performance differences originated from the underlying computational demands rather than changes in numerical implementation.

\section{RESULTS AND DISCUSSION}

\subsection{Amplitude-Driven Dynamical Operating Regimes}

From a PRC perspective, the dynamical regime is a primary factor governing computational performance, since memory generation, nonlinear state transformation, and state separability emerge directly from the underlying system dynamics. Accordingly, the excitation amplitude serves as the sole encoding channel while simultaneously acting as the bifurcation parameter governing transitions between different dynamical regimes. The metabeam response is therefore first characterized through an amplitude-driven bifurcation analysis to identify dynamically distinct operating regions before evaluating its computational performance.

Fig.~\ref{Fig_2}(a) presents representative midspan displacement histories obtained under increasing forcing amplitudes, while Fig.~\ref{Fig_2}(b) shows the corresponding bifurcation diagram. Three distinct operating regimes, periodic (P), multi-periodic (MP), and chaotic (C) are identified, providing dynamically different operating conditions for the subsequent computational benchmarks. At low forcing amplitudes, the metabeam reservoir exhibits regular oscillatory motion characterized by stable and repeatable trajectories associated with periodic dynamics. As the excitation amplitude increases, nonlinear scale engagement introduces additional oscillatory components, giving rise to multi-periodic dynamics with increasingly complex yet organized behavior. Further increase in excitation amplitude produces increasingly irregular oscillations with loss of periodic structure and broad exploration of the accessible state space, characteristic of chaotic dynamics. These transitions become more evident in the bifurcation diagram shown in Fig.~\ref{Fig_2}(b). For low excitation amplitudes, the reservoir converges to a small number of discrete response states, indicating periodic behavior and limited state diversity. As the forcing amplitude increases, successive bifurcations generate multiple coexisting oscillatory states corresponding to the multi-periodic regime. As the forcing amplitude is further increased, the response gradually evolves into a dense state distribution associated with chaotic motion, reflecting enhanced sensitivity and expansion of the accessible reservoir state manifold. These operating regimes provide distinct dynamical conditions for PRC. Their influence on nonlinear approximation, temporal prediction, and memory generation is investigated in the following.


\begin{figure}[t]
\begin{center}
\begin{tabular}{c} 
\includegraphics[width=1.0\linewidth]{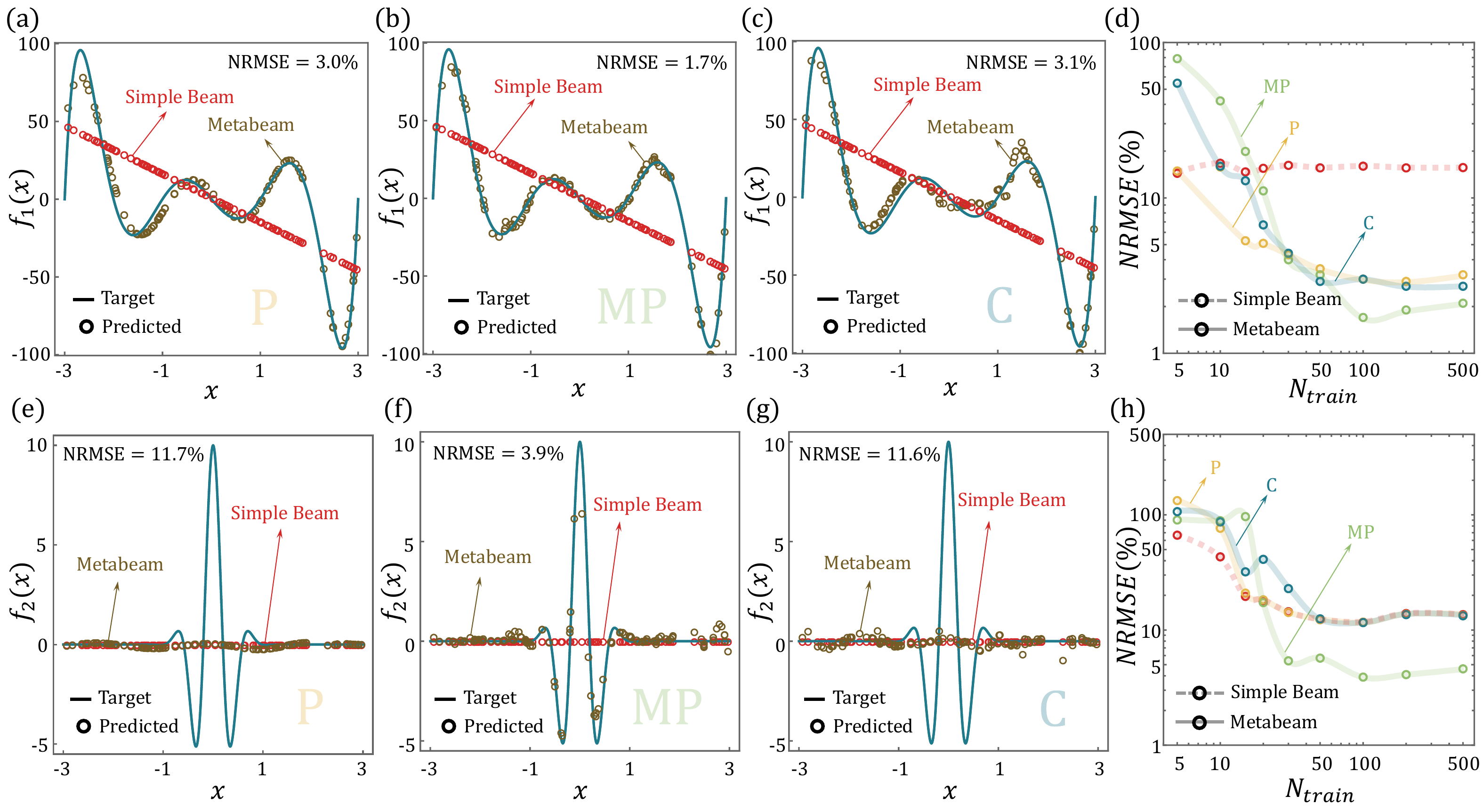}
\end{tabular}
\end{center}

\caption 
{\label{Fig_3}
{\textbf{Non-temporal nonlinear function approximation and prediction accuracy.} Predictions of the polynomial benchmark function $f_1(x)=(x-3)(x-2)(x-1)x(x+1)(x+2)(x+3)$ in the (a) periodic (P), (b) multi-periodic (MP), and (c) chaotic (C) regimes, respectively. (d) Corresponding NRMSE as a function of the training set size, ($N_{\mathrm{train}}$). Predictions of the nonlinear benchmark function $f_2(x)=10e^{-5x^2}\cos(8x)$ in the (e) P, (f) MP, and (g) C regimes, respectively. (h) Corresponding NRMSE versus $N_{\mathrm{train}}$.
}}

\end{figure} 


\subsection{Non-Temporal Nonlinear Function Approximation}

Having established dynamically distinct operating regimes for the metabeam reservoir, we first evaluate its intrinsic nonlinear representation capability independently of temporal memory requirements. To this end, non-temporal function approximation tasks are considered as controlled computational benchmarks. Unlike temporal prediction problems, these tasks do not require reconstruction of delayed dependencies and therefore provide a direct assessment of how effectively the reservoir converts low-dimensional inputs into high-dimensional nonlinear state representations.


\begin{figure}[t]
\begin{center}
\begin{tabular}{c} 
\includegraphics[width=1.0\linewidth]{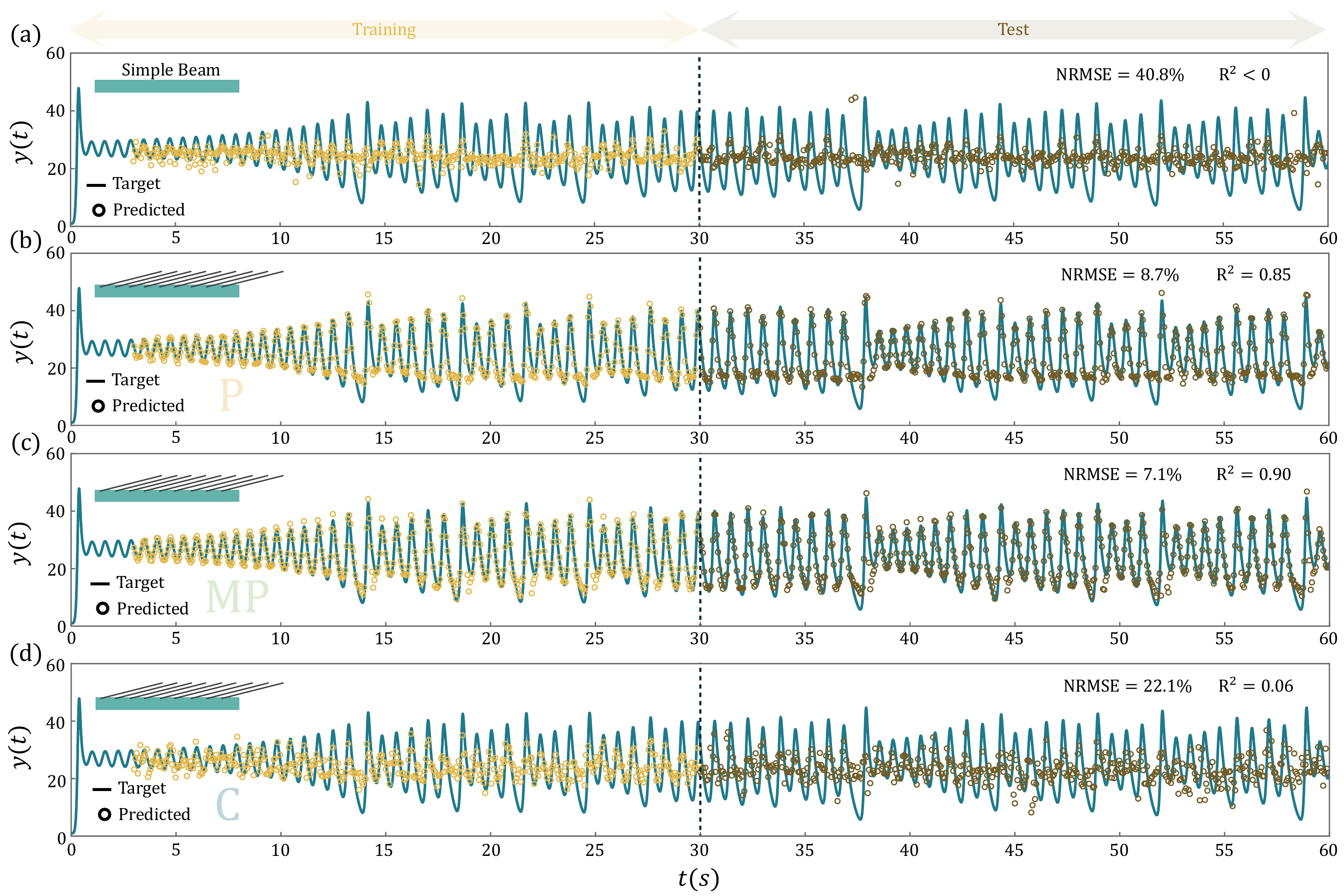}
\end{tabular}
\end{center}

\caption 
{\label{Fig_4}
{\textbf{Lorenz-63 state reconstruction across different dynamical regimes.} (a) Simple beam reservoir predictions. Lorenz-63 predictions generated by the metabeam reservoir operating in the (b) periodic (P), (c) multi-periodic (MP), and (d) chaotic (C) regimes, respectively. Solid lines denote target time series, while rings indicate reservoir predictions. The vertical dashed line separates training and test intervals.
}}

\end{figure} 


Fig.~\ref{Fig_3} presents approximation results for two benchmark functions with different structural characteristics. The first function, $f_1(x)$, is a high-order polynomial that requires accurate global high-dimensional nonlinear state representations over the entire input domain, whereas the second function, $f_2(x)$, is a localized oscillatory function requiring accurate reconstruction of rapidly varying features. For both functions, predictions are generated using reservoirs operating independently within the periodic (P), multi-periodic (MP), and chaotic (C) regimes. Prediction quality is quantified using NRMSE, while the influence of training data size is evaluated through the convergence behavior shown in Figs.~\ref{Fig_3}(d,h). 
For the polynomial function $f_1(x)$, all three operating regimes successfully capture the overall nonlinear relationship, although noticeable differences in prediction accuracy remain. Among the three regimes, the multi-periodic reservoir achieves the lowest prediction error (NRMSE = $1.7\%$), followed by the periodic ($3.0\%$) and chaotic ($3.1\%$) regimes. These results indicate that moderately complex reservoir dynamics provide the most effective nonlinear representation for this global approximation task, whereas further increases in dynamical complexity do not yield additional improvements. The corresponding convergence curves in Fig.~\ref{Fig_3}(d) further demonstrate that prediction accuracy improves steadily with increasing training data before approaching a nearly converged solution. Across all training sizes, the metabeam consistently outperforms the simple beam baseline, highlighting the beneficial role of nonlinear scale engagement in enhancing representational capability. 
A similar trend is observed for the localized oscillatory benchmark $f_2(x)$. Although all operating regimes reproduce the general structure of the target function, prediction accuracy again depends strongly on the underlying reservoir dynamics. The multi-periodic regime achieves the lowest prediction error (NRMSE = $3.9\%$), substantially outperforming both the periodic ($11.7\%$) and chaotic ($11.6\%$) regimes. This observation suggests that an intermediate level of dynamical complexity provides the most favorable balance for approximating localized nonlinear features. The convergence behavior shown in Fig.~\ref{Fig_3}(h) follows the same overall pattern, with the metabeam exhibiting faster error reduction and consistently lower prediction errors than the simple beam as the number of training samples increases. Overall, these results demonstrate that nonlinear scale engagement significantly enhances the static nonlinear representation capability of the proposed reservoir, while indicating that the multi-periodic operating regime provides the most favorable conditions for non-temporal function approximation.

\subsection{Temporal State Reconstruction of the Lorenz--63 System}

Having demonstrated the nonlinear representation capability of the metabeam reservoir, we next evaluate its ability to process continuously evolving temporal dynamics. To this end, the Lorenz–63 benchmark is considered as a state reconstruction task in which the reservoir receives only partial observations of an underlying chaotic system. Unlike the non-temporal benchmarks discussed previously, successful prediction requires both retention of temporal information and reconstruction of hidden nonlinear relationships embedded within the observed trajectory. Specifically, the observable Lorenz coordinate $x(t)$ is provided as input to the reservoir, while the corresponding hidden state $z(t)$ is reconstructed through the trained readout. 

Fig.~\ref{Fig_4} compares representative predictions obtained using both the simple beam baseline and the metabeam reservoir operating within the periodic (P), multi-periodic (MP), and chaotic (C) regimes. The simple beam exhibits limited reconstruction capability and produces noticeable deviations from the target trajectory throughout both training and testing intervals. In contrast, the metabeam reservoir substantially improves prediction accuracy across all operating regimes, indicating that contact-induced nonlinear dynamics generate richer reservoir representations for continuous-state inference. Among the three regimes, the multi-periodic reservoir achieves the lowest prediction error and provides the closest agreement with the target trajectory, achieving an NRMSE of approximately $7.1\%$. The periodic regime also demonstrates accurate reconstruction, with an NRMSE of approximately $8.7\%$, whereas the chaotic regime exhibits substantially larger deviations and reduced long-term agreement, yielding an NRMSE of approximately $22.1\%$. In contrast, the simple linear beam fails to reconstruct the underlying dynamics, resulting in an NRMSE of approximately $40.8\%$. These results indicate that the highest prediction accuracy is achieved within the multi-periodic regime, suggesting that an intermediate level of dynamical complexity provides the most favorable balance between rich nonlinear state representations and stable temporal evolution for continuous-state reconstruction. This observation is consistent with previous studies showing that reservoir performance is governed by a trade-off between high-dimensional nonlinear state representations and fading memory, rather than by maximizing dynamical complexity alone \cite{inubushi2017reservoir,he2025role}. Excessively chaotic dynamics enrich nonlinear representations but simultaneously reduce temporal coherence and memory retention, whereas purely periodic dynamics preserve memory at the expense of representational richness. The superior performance observed in the multi-periodic regime therefore indicates that an intermediate dynamical state provides the most effective balance between these competing computational requirements.


\begin{figure}[t]
\begin{center}
\begin{tabular}{c} 
\includegraphics[width=1.0\linewidth]{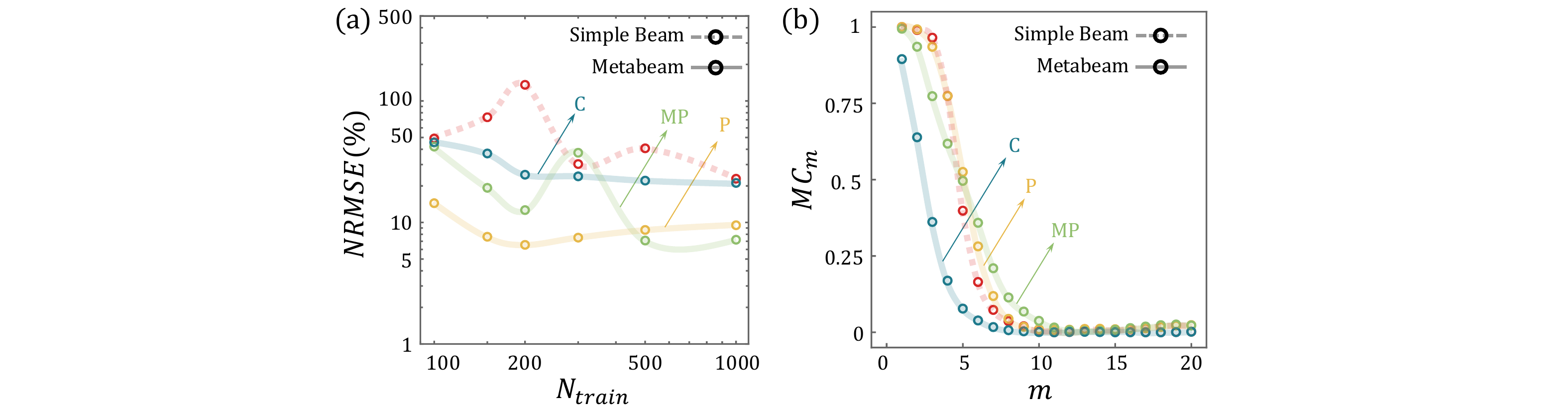}
\end{tabular}
\end{center}

\caption 
{\label{Fig_5}
{\textbf{Prediction accuracy and memory characteristics for the Lorenz-63 task} (a) NRMSE as a function of the training set size, $N_{\mathrm{train}}$, for the simple beam and metabeam reservoirs operating in the periodic (P), multi-periodic (MP), and chaotic (C) regimes. (b) Memory capacity ($MC_m$) as a function of memory depth ($m$) for the corresponding reservoirs.
}}

\end{figure} 


Fig.~\ref{Fig_5}(a,b) provides additional insight into this behavior by summarizing prediction accuracy as a function of training set size together with the corresponding memory characteristics. Prediction accuracy improves systematically as additional training states become available, although convergence behavior differs substantially across operating regimes. The periodic reservoir converges rapidly toward low prediction error. These computational trends are consistent with the measured memory characteristics shown in Fig.~\ref{Fig_5}(b). The periodic reservoir exhibits the largest memory capacity ($MC=4.84$), followed by the multi-periodic ($MC=4.75$), simple beam ($MC=4.44$), and chaotic ($MC=2.22$) configurations. Reservoirs exhibiting stronger short-term memory retention generally achieve superior reconstruction performance. Although chaotic dynamics increase state diversity, excessive dynamical complexity reduces effective memory utilization and limits prediction quality. These results suggest that continuous chaotic state reconstruction benefits primarily from stable fading memory rather than maximal nonlinear expansion. Overall, the Lorenz–63 benchmark demonstrates that the proposed metabeam reservoir can infer hidden states of a continuous chaotic system using only partial observations. More importantly, the results indicate that reservoir performance is governed by a balance between high-dimensional nonlinear state representations and temporal memory, motivating the subsequent investigation of increasingly demanding temporal prediction tasks through the NARMA benchmarks.


\begin{figure}[t]
\begin{center}
\begin{tabular}{c} 
\includegraphics[width=1.0\linewidth]{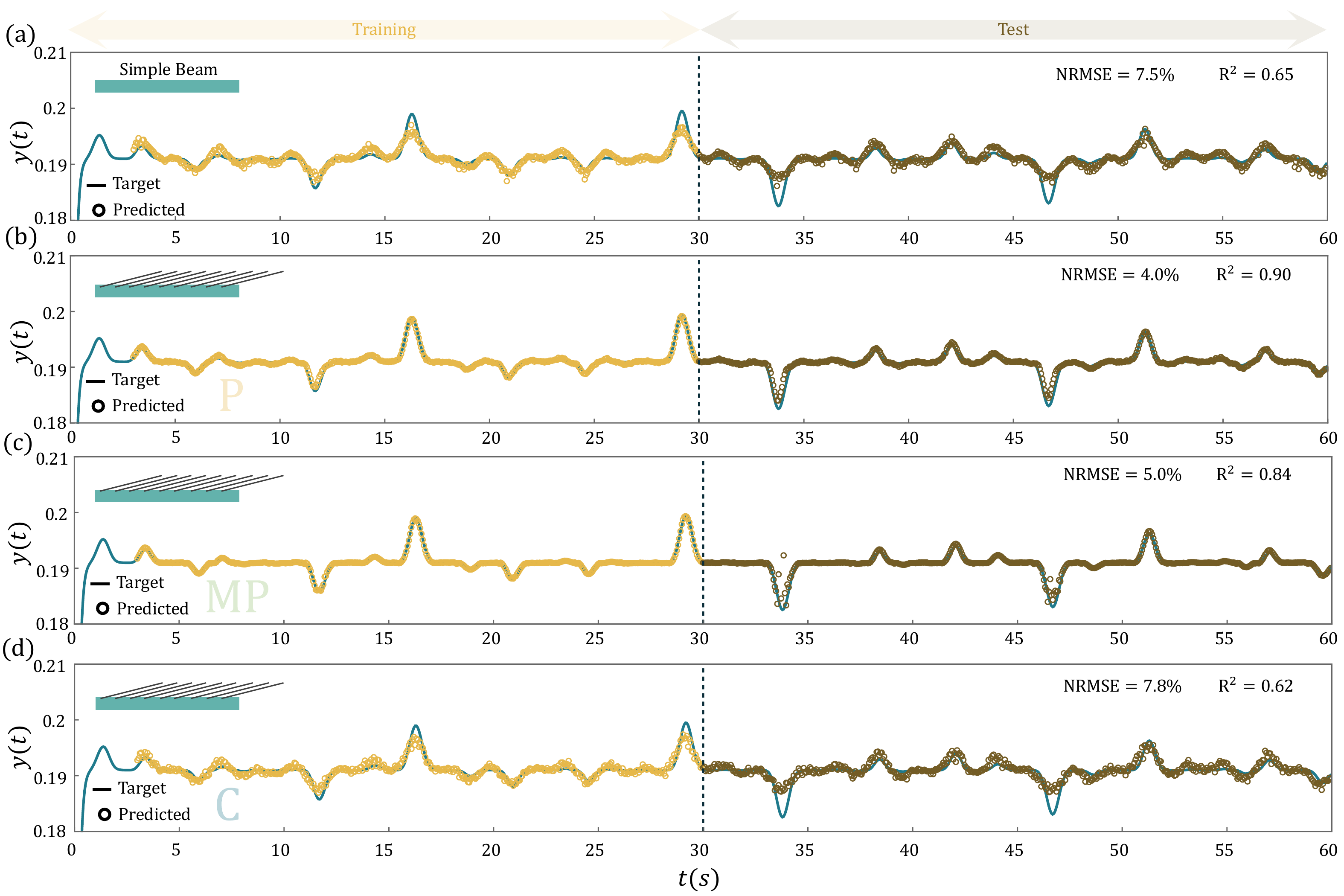}
\end{tabular}
\end{center}

\caption 
{\label{Fig_6}
{\textbf{NARMA-2 prediction performance.} (a) Predictions obtained using a simple beam reservoir. NARMA-2 predictions generated by the metabeam reservoir operating in the (b) periodic (P), (c) multi-periodic (MP), and (d) chaotic (C) regimes, respectively. Solid lines denote target outputs, while rings represent reservoir predictions. The vertical dashed line separates training and test intervals. 
}}

\end{figure} 



\begin{figure}[t]
\begin{center}
\begin{tabular}{c} 
\includegraphics[width=1.0\linewidth]{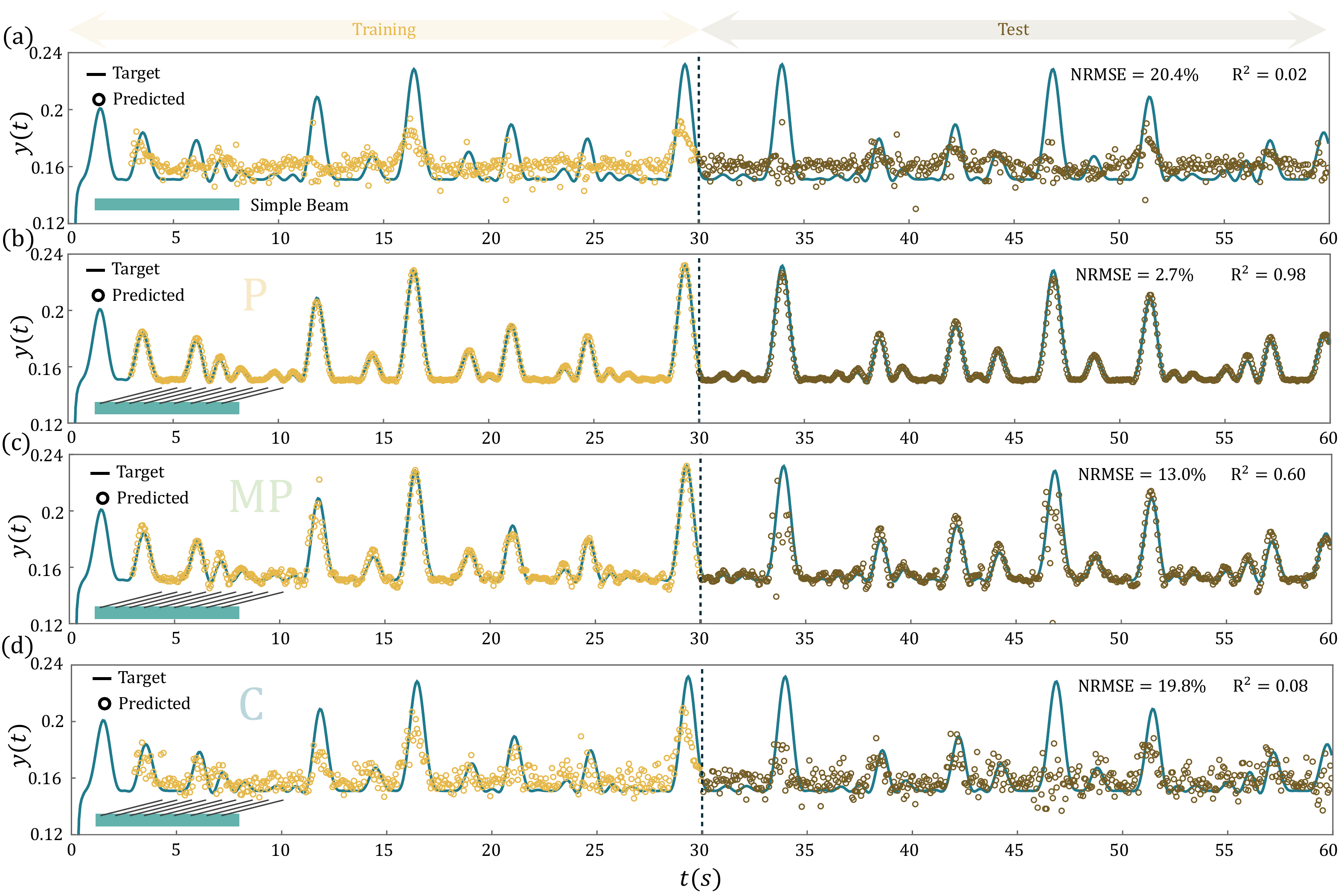}
\end{tabular}
\end{center}

\caption 
{\label{Fig_7}
{\textbf{NARMA-5 prediction performance.} (a) Prediction results for the simple beam reservoir. NARMA-5 predictions generated by the metabeam reservoir operating in the (b) periodic (P), (c) multi-periodic (MP), and (d) chaotic (C) regimes, respectively. Solid lines denote target outputs, while rings represent reservoir predictions. The vertical dashed line separates training and test intervals.
}}

\end{figure} 


\subsection{Temporal Processing Using NARMA Benchmarks}

To systematically evaluate the temporal processing capability of the proposed metabeam reservoir, nonlinear autoregressive moving average (NARMA) benchmarks with progressively increasing temporal order are considered. Unlike the Lorenz–63 reconstruction task, where prediction relies on extracting hidden relationships from a continuous trajectory, NARMA benchmarks explicitly require reconstruction of delayed nonlinear interactions over prescribed temporal horizons. As the benchmark order increases, successful prediction becomes increasingly dependent on effective memory utilization, since target outputs depend simultaneously on present inputs, nonlinear state interactions, and delayed information extending across previous states. Accordingly, NARMA tasks provide a controlled framework for examining the balance between high-dimensional nonlinear state representations and temporal information retention in the proposed reservoir.

Fig.~\ref{Fig_6} presents prediction results for the NARMA-2 benchmark, which imposes relatively modest memory requirements and therefore evaluates the reservoir's nonlinear representation capability. All metabeam reservoirs outperform the simple beam baseline, demonstrating that contact-induced nonlinear dynamics provide richer computational states than linear structural dynamics alone. Among the examined operating regimes, the periodic reservoir achieves the closest agreement with the target trajectory, yielding the lowest prediction error (NRMSE = $4.0\%$). The multi-periodic reservoir also demonstrates excellent predictive accuracy, with only a modest increase in error (NRMSE = $5.0\%$), whereas the chaotic reservoir exhibits noticeably larger deviations (NRMSE = $7.8\%$). These results indicate that under modest memory requirements, stable reservoir dynamics provide more effective nonlinear state representations than highly irregular chaotic dynamics, while the multi-periodic regime remains competitive with only a slight reduction in accuracy. 

Increasing the benchmark complexity to NARMA-5 introduces stronger dependence on delayed nonlinear interactions and therefore places greater demand on reservoir memory. As shown in Fig.~\ref{Fig_7}, prediction accuracy decreases relative to NARMA-2, reflecting the increased difficulty associated with longer temporal dependencies. The influence of the reservoir operating regime also becomes substantially more pronounced. The periodic reservoir maintains excellent predictive performance and achieves the lowest prediction error (NRMSE = $2.7\%$), demonstrating that stable reservoir dynamics remain highly effective under moderate memory demands. In contrast, the multi-periodic regime exhibits a noticeable reduction in accuracy (NRMSE = $13.0\%$), while the chaotic regime undergoes severe performance degradation (NRMSE = $19.8\%$), approaching the performance of the simple beam baseline (NRMSE = $20.4\%$). These results suggest that increasing temporal memory requirements progressively favor stable reservoir dynamics, whereas excessive dynamical complexity degrades the reservoir’s ability to preserve and exploit long-range temporal information.

\begin{figure}[t]
\begin{center}
\begin{tabular}{c} 
\includegraphics[width=1.0\linewidth]{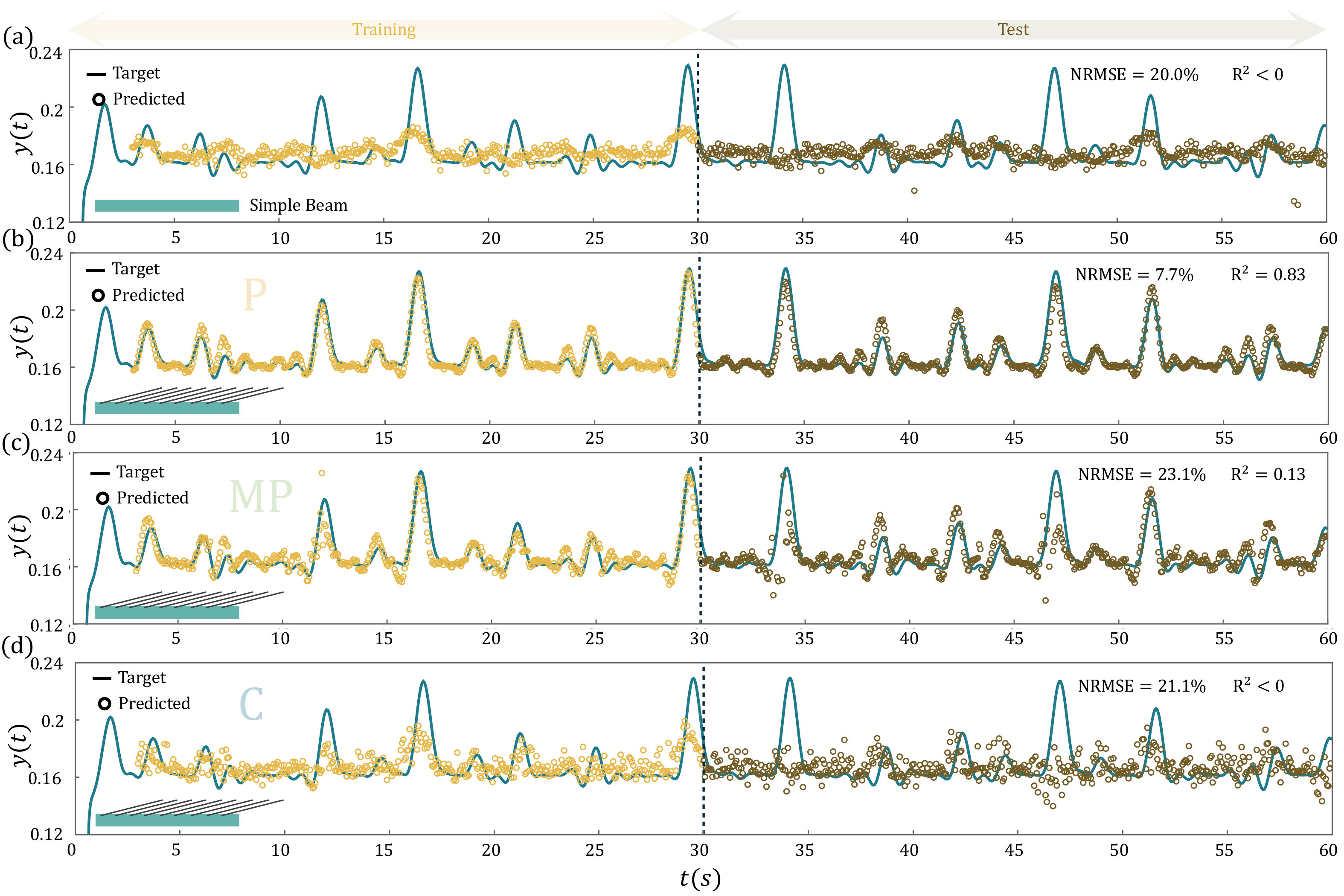}
\end{tabular}
\end{center}

\caption 
{\label{Fig_8}
{\textbf{NARMA-10 prediction performance.} (a) Simple beam reservoir prediction. NARMA-10 predictions generated by the metabeam reservoir operating in the (b) periodic (P), (c) multi-periodic (MP), and (d) chaotic (C) regimes, respectively. Solid lines denote target outputs, while rings represent reservoir predictions. The vertical dashed line separates training and test intervals.
}}

\end{figure}

The NARMA-10 benchmark represents the most demanding temporal prediction task considered in this study and requires simultaneous reconstruction of extended delayed interactions and nonlinear coupling across multiple excitation intervals. As illustrated in Fig.~\ref{Fig_8}, prediction difficulty increases substantially relative to the lower-order benchmarks, leading to a noticeable reduction in prediction accuracy across all operating regimes. Nevertheless, clear differences in computational capability persist among the examined dynamical regimes. The periodic reservoir continues to provide the closest agreement with the target sequence, achieving the lowest prediction error (NRMSE = $7.7\%$). The multi-periodic regime remains capable of capturing the dominant temporal behavior but exhibits increased deviations (NRMSE = $13.1\%$), whereas the chaotic regime undergoes substantial performance degradation (NRMSE = $21.1\%$), yielding performance comparable to the simple beam baseline (NRMSE = $20.0\%$). These results indicate that under demanding long-memory tasks, stable reservoir dynamics provide more effective exploitation of temporal information than increasingly irregular dynamical evolution, while excessive dynamical complexity offers little computational advantage.


\begin{figure}[t]
\begin{center}
\begin{tabular}{c} 
\includegraphics[width=1.0\linewidth]{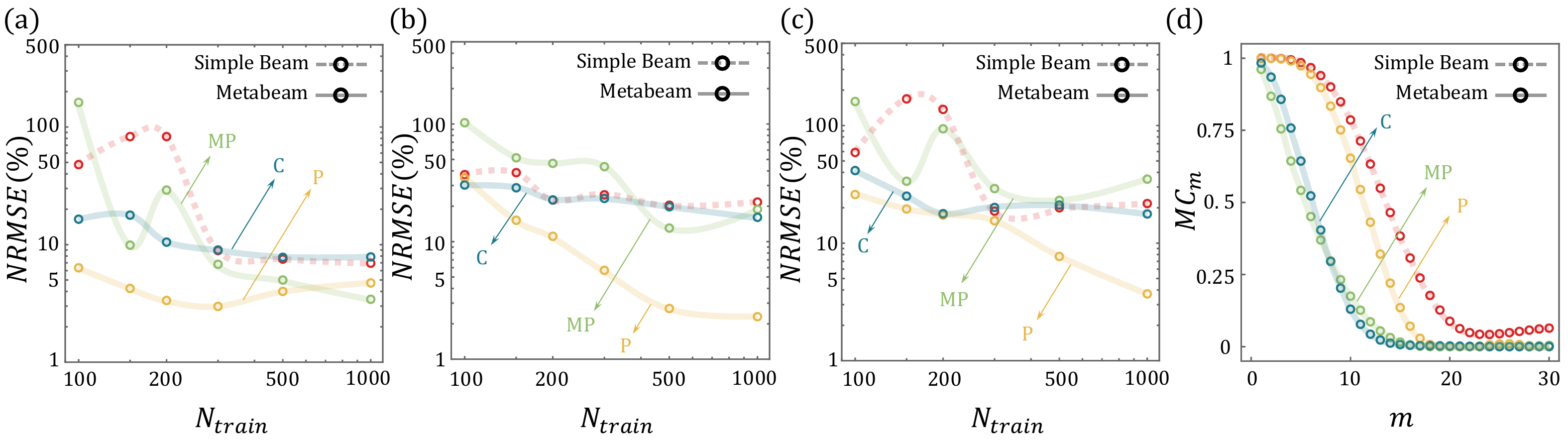}
\end{tabular}
\end{center}
\caption 
{\label{Fig_9}
{\textbf{Prediction accuracy and memory characteristics across the NARMA benchmarks.} NRMSE versus training set size, $N_{\mathrm{train}}$, for the (a) NARMA-2, (b) NARMA-5, and (c) NARMA-10, respectively. Results are shown for the simple beam and metabeam reservoirs operating in the periodic (P), multi-periodic (MP), and chaotic (C) regimes. (d) Corresponding memory capacity ($MC_m$) as a function of memory depth ($m$). 
}}
\end{figure} 


Fig.~\ref{Fig_9} summarizes the prediction accuracy and memory characteristics across all NARMA benchmarks. Figs.~\ref{Fig_9}(a–c) show that, for all benchmark orders, increasing the size of the training dataset generally reduces the prediction error; however, the convergence behavior depends strongly on the underlying dynamical regime. For the NARMA-2 benchmark, all metabeam configurations outperform the simple beam baseline, with the periodic reservoir consistently achieving the lowest prediction error. As the benchmark order increases from NARMA-2 to NARMA-5 and NARMA-10, the performance gap between the operating regimes becomes progressively more pronounced. The periodic reservoir maintains the fastest convergence and the lowest final NRMSE, the multi-periodic reservoir exhibits intermediate performance, and the chaotic reservoir gradually approaches the prediction accuracy of the simple beam baseline. Fig.~\ref{Fig_9}(d) further illustrates the corresponding memory characteristics. The simple beam exhibits the largest memory capacity ($MC=13.63$), followed by the periodic ($MC=10.84$), chaotic ($MC=5.92$), and multi-periodic ($MC=5.63$) reservoirs. Despite the simple beam exhibiting the largest memory capacity, its prediction accuracy remains consistently inferior to that of the metabeam operating in the periodic regime, indicating that memory alone is insufficient for optimal computation. Instead, the periodic regime provides a more effective balance between high-dimensional nonlinear representation and temporal information retention, leading to superior performance across all NARMA benchmarks. The substantially lower memory capacities of the multi-periodic and chaotic regimes are accompanied by progressively degraded prediction accuracy, further demonstrating that effective reservoir computation requires an appropriate balance between memory and nonlinear dynamics rather than maximizing either property alone.

\section{CONCLUSION}
This study introduces a geometry-programmable metabeam inspired by hierarchical overlapping fish-scale architectures as a platform for enabling physical reservoir computing (PRC). Contact interactions between embedded scales enable the structure to access periodic, multi-periodic, and chaotic dynamical regimes, thereby generating reservoir states with different levels of nonlinear richness and temporal dependence. The computational capability of the proposed metabeam was evaluated through static nonlinear function approximation, Lorenz-63 prediction, and NARMA benchmark tasks, using an equivalent simple beam as a linear baseline. Across the considered tasks, the metabeam generally outperformed the linear beam by providing more informative state representations, lower prediction errors, improved temporal-memory characteristics, and reduced training-data requirements. The results also demonstrated that computational performance depends strongly on the operating regime and does not necessarily improve with increasing dynamical complexity. Periodic and multi-periodic responses generally provided the most favorable computational conditions, whereas chaotic dynamics produced larger prediction errors in the memory-dependent tasks. In particular, the multi-periodic regime achieved the highest accuracy for Lorenz-63 prediction, while the periodic regime provided the strongest overall performance in the NARMA tasks. These findings indicate that effective physical reservoir computation emerges from an appropriate balance between nonlinear representational capability, temporal-memory retention, and dynamical stability rather than from maximally complex or chaotic behavior. The observed computational gains arise solely from geometry-induced contact nonlinearities, without requiring active materials, embedded electronic components, or modification of the intrinsic reservoir parameters during operation. Information is introduced through forcing-amplitude modulation, while the physical configuration of the metabeam remains unchanged. The results therefore establish structural geometry and contact interactions as programmable resources for physically embodied information processing. More broadly, the proposed framework extends architected mechanical systems beyond conventional load-bearing and vibration-control functions and provides a foundation for the development of mechanically intelligent structures for sensing, adaptive robotics, and embodied computing.

\acknowledgments 
This work was supported by the U.S. National Science Foundation (NSF) under the Civil, Mechanical, and Manufacturing Innovation (CMMI) CAREER Awards No. 1943886 and No. 2028338.

\bibliography{main} 

@article{he2025role,
  title={The role of nonlinearity, dimensionality, memory, and input on a mechanical oscillator reservoir computer},
  author={He, Shan and Musgrave, Patrick F},
  journal={Neurocomputing},
  volume={652},
  pages={131158},
  year={2025},
  publisher={Elsevier}
}

@article{mandal2022machine,
  title={Machine-learning potential of a single pendulum},
  author={Mandal, Swarnendu and Sinha, Sudeshna and Shrimali, Manish Dev},
  journal={Physical Review E},
  volume={105},
  number={5},
  pages={054203},
  year={2022},
  publisher={APS}
}

@article{ghosh2014contact,
  title={Contact kinematics of biomimetic scales},
  author={Ghosh, Ranajay and Ebrahimi, Hamid and Vaziri, Ashkan},
  journal={Applied Physics Letters},
  volume={105},
  number={23},
  year={2014},
  publisher={AIP Publishing}
}

@article{jaeger2004harnessing,
  title={Harnessing nonlinearity: Predicting chaotic systems and saving energy in wireless communication},
  author={Jaeger, Herbert and Haas, Harald},
  journal={science},
  volume={304},
  number={5667},
  pages={78--80},
  year={2004},
  publisher={American Association for the Advancement of Science}
}

@article{cucchi2022hands,
  title={Hands-on reservoir computing: a tutorial for practical implementation},
  author={Cucchi, Matteo and Abreu, Steven and Ciccone, Giuseppe and Brunner, Daniel and Kleemann, Hans},
  journal={Neuromorphic Computing and Engineering},
  volume={2},
  number={3},
  pages={032002},
  year={2022},
  publisher={IoP Publishing}
}

@inproceedings{bateniparvar2024dynamic,
  title={Dynamic excitations in periodic fish scale inspired structures},
  author={Bateniparvar, Omid and Hossain, Md Shahjahan and Ghosh, Ranajay},
  booktitle={Bioinspiration, Biomimetics, and Bioreplication XIV},
  volume={12944},
  pages={17--23},
  year={2024},
  organization={SPIE}
}

@article{tanaka2019recent,
  title={Recent advances in physical reservoir computing: A review},
  author={Tanaka, Gouhei and Yamane, Toshiyuki and H{\'e}roux, Jean Benoit and Nakane, Ryosho and Kanazawa, Naoki and Takeda, Seiji and Numata, Hidetoshi and Nakano, Daiju and Hirose, Akira},
  journal={Neural Networks},
  volume={115},
  pages={100--123},
  year={2019},
  publisher={Elsevier}
}

@article{maass2002real,
  title={Real-time computing without stable states: A new framework for neural computation based on perturbations},
  author={Maass, Wolfgang and Natschl{\"a}ger, Thomas and Markram, Henry},
  journal={Neural computation},
  volume={14},
  number={11},
  pages={2531--2560},
  year={2002},
  publisher={MIT Press}
}

@article{jaeger2001echo,
  title={The “echo state” approach to analysing and training recurrent neural networks-with an erratum note},
  author={Jaeger, Herbert},
  journal={Bonn, Germany: German national research center for information technology gmd technical report},
  volume={148},
  number={34},
  pages={13},
  year={2001},
  publisher={Bonn}
}

@article{nakajima2020physical,
  title={Physical reservoir computing—an introductory perspective},
  author={Nakajima, Kohei},
  journal={Japanese Journal of Applied Physics},
  volume={59},
  number={6},
  pages={060501},
  year={2020},
  publisher={IOP Publishing}
}

@article{dambre2012information,
  title={Information processing capacity of dynamical systems},
  author={Dambre, Joni and Verstraeten, David and Schrauwen, Benjamin and Massar, Serge},
  journal={Scientific reports},
  volume={2},
  number={1},
  pages={514},
  year={2012},
  publisher={Nature Publishing Group UK London}
}

@article{gallicchio2017echo,
  title={Echo state property of deep reservoir computing networks},
  author={Gallicchio, Claudio and Micheli, Alessio},
  journal={Cognitive Computation},
  volume={9},
  number={3},
  pages={337--350},
  year={2017},
  publisher={Springer}
}

@article{terajima2025multifunctional,
  title={Multifunctional physical reservoir computing in soft tensegrity robots},
  author={Terajima, Ryo and Inoue, Katsuma and Nakajima, Kohei and Kuniyoshi, Yasuo},
  journal={Chaos: An Interdisciplinary Journal of Nonlinear Science},
  volume={35},
  number={8},
  year={2025},
  publisher={AIP Publishing}
}

@article{caluwaerts2013locomotion,
  title={Locomotion without a brain: physical reservoir computing in tensegrity structures},
  author={Caluwaerts, Ken and D'Haene, Michiel and Verstraeten, David and Schrauwen, Benjamin},
  journal={Artificial life},
  volume={19},
  number={1},
  pages={35--66},
  year={2013},
  publisher={MIT Press One Rogers Street, Cambridge, MA 02142-1209, USA journals-info~…}
}

@article{austin2026physical,
  title={Physical Parameter Dependencies in Mechanical Reservoir Computing: Structural Analysis, Actuation, and Improved Processing},
  author={Austin, Max and Nakajima, Kohei},
  journal={IEEE Transactions on Neural Networks and Learning Systems},
  year={2026},
  publisher={IEEE}
}

@article{dion2018reservoir,
  title={Reservoir computing with a single delay-coupled non-linear mechanical oscillator},
  author={Dion, Guillaume and Mejaouri, Salim and Sylvestre, Julien},
  journal={Journal of Applied Physics},
  volume={124},
  number={15},
  year={2018},
  publisher={AIP Publishing}
}

@article{furuta2018macromagnetic,
  title={Macromagnetic simulation for reservoir computing utilizing spin dynamics in magnetic tunnel junctions},
  author={Furuta, Taishi and Fujii, Keisuke and Nakajima, Kohei and Tsunegi, Sumito and Kubota, Hitoshi and Suzuki, Yoshishige and Miwa, Shinji},
  journal={Physical Review Applied},
  volume={10},
  number={3},
  pages={034063},
  year={2018},
  publisher={APS}
}

@article{nakajima2015information,
  title={Information processing via physical soft body},
  author={Nakajima, Kohei and Hauser, Helmut and Li, Tao and Pfeifer, Rolf},
  journal={Scientific reports},
  volume={5},
  number={1},
  pages={10487},
  year={2015},
  publisher={Nature Publishing Group UK London}
}

@article{bateniparvar2026chaotic,
  title={Chaotic Flexural Vibrations in Biomimetic Scale Substrates},
  author={Bateniparvar, Omid and Farahmand, Farzan and Ghosh, Ranajay},
  journal={arXiv preprint arXiv:2604.13329},
  year={2026}
}

@article{meyers2008biological,
  title={Biological materials: structure and mechanical properties},
  author={Meyers, Marc Andr{\'e} and Chen, Po-Yu and Lin, Albert Yu-Min and Seki, Yasuaki},
  journal={Progress in materials science},
  volume={53},
  number={1},
  pages={1--206},
  year={2008},
  publisher={Elsevier}
}

@article{fratzl2007nature,
  title={Nature’s hierarchical materials},
  author={Fratzl, Peter and Weinkamer, Richard},
  journal={Progress in materials Science},
  volume={52},
  number={8},
  pages={1263--1334},
  year={2007},
  publisher={Elsevier}
}

@article{vincent2006biomimetics,
  title={Biomimetics: its practice and theory},
  author={Vincent, Julian FV and Bogatyreva, Olga A and Bogatyrev, Nikolaj R and Bowyer, Adrian and Pahl, Anja-Karina},
  journal={Journal of the Royal Society Interface},
  volume={3},
  number={9},
  pages={471--482},
  year={2006},
  publisher={The Royal Society}
}

@article{hossain2022fish,
  title={Fish scale inspired structures—A review of materials, manufacturing and models},
  author={Hossain, Md Shahjahan and Ebrahimi, Hossein and Ghosh, Ranajay},
  journal={Bioinspiration \& Biomimetics},
  volume={17},
  number={6},
  pages={061001},
  year={2022},
  publisher={IOP Publishing}
}

@article{ebrahimi2023material,
  title={Material-geometry interplay in damping of biomimetic scale beams},
  author={Ebrahimi, Hossein and Krsmanovic, Milos and Ali, Hessein and Ghosh, Ranajay},
  journal={Applied Physics Letters},
  volume={123},
  number={8},
  year={2023},
  publisher={AIP Publishing}
}

@article{ali2019frictional,
  title={Frictional damping from biomimetic scales},
  author={Ali, Hessein and Ebrahimi, Hossein and Ghosh, Ranajay},
  journal={Scientific reports},
  volume={9},
  number={1},
  pages={14628},
  year={2019},
  publisher={Nature Publishing Group UK London}
}

@article{sarkar2025bending,
  title={Bending mechanics of biomimetic scale plates},
  author={Sarkar, Pranta Rahman and Ebrahimi, Hossein and Hossain, Md Shahjahan and Ali, Hessein and Ghosh, Ranajay},
  journal={European Journal of Mechanics-A/Solids},
  volume={113},
  pages={105664},
  year={2025},
  publisher={Elsevier}
}

@article{duport2012all,
  title={All-optical reservoir computing},
  author={Duport, Fran{\c{c}}ois and Schneider, Bendix and Smerieri, Anteo and Haelterman, Marc and Massar, Serge},
  journal={Optics express},
  volume={20},
  number={20},
  pages={22783--22795},
  year={2012},
  publisher={Optical Society of America}
}

@article{denis2018all,
  title={All-optical reservoir computing on a photonic chip using silicon-based ring resonators},
  author={Denis-Le Coarer, Florian and Sciamanna, Marc and Katumba, Andrew and Freiberger, Matthias and Dambre, Joni and Bienstman, Peter and Rontani, Damien},
  journal={IEEE Journal of Selected Topics in Quantum Electronics},
  volume={24},
  number={6},
  pages={1--8},
  year={2018},
  publisher={IEEE}
}

@article{liang2024physical,
  title={Physical reservoir computing with emerging electronics},
  author={Liang, Xiangpeng and Tang, Jianshi and Zhong, Yanan and Gao, Bin and Qian, He and Wu, Huaqiang},
  journal={Nature Electronics},
  volume={7},
  number={3},
  pages={193--206},
  year={2024},
  publisher={Nature Publishing Group UK London}
}

@inproceedings{nowshin2020recent,
  title={Recent advances in reservoir computing with a focus on electronic reservoirs},
  author={Nowshin, Fabiha and Zhang, Yuhao and Liu, Lingjia and Yi, Yang},
  booktitle={2020 11th International Green and Sustainable Computing Workshops (IGSC)},
  pages={1--8},
  year={2020},
  organization={IEEE}
}

@inproceedings{fernando2003pattern,
  title={Pattern recognition in a bucket},
  author={Fernando, Chrisantha and Sojakka, Sampsa},
  booktitle={European conference on artificial life},
  pages={588--597},
  year={2003},
  organization={Springer}
}

@article{heuthe2026reservoir,
  title={Reservoir computing from collective dynamics of active colloidal oscillators},
  author={Heuthe, Veit-Lorenz and Seemann, Lukas and Tovey, Samuel and Bechinger, Clemens},
  journal={Communications AI \& Computing},
  volume={1},
  number={1},
  pages={6},
  year={2026},
  publisher={Nature Publishing Group UK London}
}

@article{coulombe2017computing,
  title={Computing with networks of nonlinear mechanical oscillators},
  author={Coulombe, Jean C and York, Mark CA and Sylvestre, Julien},
  journal={PloS one},
  volume={12},
  number={6},
  pages={e0178663},
  year={2017},
  publisher={Public Library of Science San Francisco, CA USA}
}

@article{shirmohammadli2024physics,
  title={Physics-based approach to developing physical reservoir computers},
  author={Shirmohammadli, Vahideh and Bahreyni, Behraad},
  journal={Physical Review Research},
  volume={6},
  number={3},
  pages={033055},
  year={2024},
  publisher={APS}
}

@article{gorb2008biological,
  title={Biological attachment devices: exploring nature's diversity for biomimetics},
  author={Gorb, Stanislav N},
  journal={Philosophical Transactions of the Royal Society A: Mathematical, Physical and Engineering Sciences},
  volume={366},
  number={1870},
  pages={1557--1574},
  year={2008},
  publisher={The Royal Society London}
}

@article{noel2018tongue,
  title={The tongue as a gripper},
  author={Noel, Alexis C and Hu, David L},
  journal={Journal of Experimental Biology},
  volume={221},
  number={7},
  pages={jeb176289},
  year={2018},
  publisher={The Company of Biologists Ltd}
}

@article{chuong2000evo,
  title={Evo-Devo of feathers and scales: Building complex epithelial appendages: commentary},
  author={Chuong, Cheng-Ming and Chodankar, Rajas and Widelitz, Randall B and Jiang, Ting-Xin},
  journal={Current opinion in genetics \& development},
  volume={10},
  number={4},
  pages={449--456},
  year={2000},
  publisher={Elsevier}
}

@article{krsmanovic2023flutter,
  title={Fur flutter in fluid flow fends off foulers},
  author={Krsmanovic, Milos and Ghosh, Ranajay and Dickerson, Andrew K},
  journal={Journal of the Royal Society Interface},
  volume={20},
  number={209},
  year={2023},
  publisher={The Royal Society}
}

@incollection{lorenz2017deterministic,
  title={Deterministic nonperiodic flow 1},
  author={Lorenz, Edward N},
  booktitle={Universality in Chaos, 2nd edition},
  pages={367--378},
  year={2017},
  publisher={Routledge}
}

@article{atiya2000new,
  title={New results on recurrent network training: unifying the algorithms and accelerating convergence},
  author={Atiya, Amir F and Parlos, Alexander G},
  journal={IEEE transactions on neural networks},
  volume={11},
  number={3},
  pages={697--709},
  year={2000},
  publisher={IEEE}
}

@article{gauthier2021next,
  title={Next generation reservoir computing},
  author={Gauthier, Daniel J and Bollt, Erik and Griffith, Aaron and Barbosa, Wendson AS},
  journal={Nature communications},
  volume={12},
  number={1},
  pages={5564},
  year={2021},
  publisher={Nature Publishing Group UK London}
}

@article{arun2024reservoir,
  title={Reservoir computing with logistic map},
  author={Arun, R and Sathish Aravindh, M and Venkatesan, A and Lakshmanan, M},
  journal={Physical Review E},
  volume={110},
  number={3},
  pages={034204},
  year={2024},
  publisher={APS}
}

@incollection{lukovsevivcius2012practical,
  title={A practical guide to applying echo state networks},
  author={Luko{\v{s}}evi{\v{c}}ius, Mantas},
  booktitle={Neural Networks: Tricks of the Trade: Second Edition},
  pages={659--686},
  year={2012},
  publisher={Springer}
}

@article{hoerl2000ridge,
  title={Ridge regression: Biased estimation for nonorthogonal problems},
  author={Hoerl, Arthur E and Kennard, Robert W},
  journal={Technometrics},
  volume={42},
  number={1},
  pages={80--86},
  year={2000},
  publisher={Taylor \& Francis}
}

@inproceedings{penrose1955generalized,
  title={A generalized inverse for matrices},
  author={Penrose, Roger},
  booktitle={Mathematical proceedings of the Cambridge philosophical society},
  volume={51},
  number={3},
  pages={406--413},
  year={1955},
  organization={Cambridge University Press}
}

@article{jaeger2001short,
  title={Short term memory in echo state networks},
  author={Jaeger, Herbert},
  year={2001},
  publisher={GMD Forschungszentrum Informationstechnik}
}

@article{stepney2024physical,
  title={Physical reservoir computing: a tutorial},
  author={Stepney, Susan},
  journal={Natural Computing},
  volume={23},
  number={4},
  pages={665--685},
  year={2024},
  publisher={Springer}
}

@article{wringe2025reservoir,
  title={Reservoir computing benchmarks: a tutorial review and critique},
  author={Wringe, Chester and Trefzer, Martin and Stepney, Susan},
  journal={International Journal of Parallel, Emergent and Distributed Systems},
  volume={40},
  number={4},
  pages={313--351},
  year={2025},
  publisher={Taylor \& Francis}
}

@article{kiyabu2026geometric,
  title={Geometric nonlinearity for 2D mechanical reservoir computing},
  author={Kiyabu, Steven and Toledo, Sebastian and He, Shan and Criner, Amanda and Musgrave, Patrick and Buskohl, Philip R},
  journal={Journal of Intelligent Material Systems and Structures},
  pages={1045389X261420480},
  year={2026},
  publisher={SAGE Publications Sage UK: London, England}
}

@article{he2026embodying,
  title={Embodying Intelligence into Mechanical Metamaterials via Reservoir Computing},
  author={He, Shan and Kiyabu, Steven and Buskohl, Philip R and Musgrave, Patrick},
  journal={arXiv preprint arXiv:2605.19098},
  year={2026}
}

@inproceedings{bateniparvar2026biomimetic,
  title={A biomimetic scale substrate as a physical reservoir computer},
  author={Bateniparvar, Omid and Ghosh, Ranajay},
  booktitle={Biologically Inspired Materials, Processes, and Systems (BIMPS) 2026},
  volume={13944},
  pages={40--50},
  year={2026},
  organization={SPIE}
}

@article{bhovad2021physical,
  title={Physical reservoir computing with origami and its application to robotic crawling},
  author={Bhovad, Priyanka and Li, Suyi},
  journal={Scientific Reports},
  volume={11},
  number={1},
  pages={13002},
  year={2021},
  publisher={Nature Publishing Group UK London}
}

@article{wang2023building,
  title={Building intelligence in the mechanical domain—Harvesting the reservoir computing power in origami to achieve information perception tasks},
  author={Wang, Jun and Li, Suyi},
  journal={Advanced Intelligent Systems},
  volume={5},
  number={9},
  pages={2300086},
  year={2023},
  publisher={Wiley Online Library}
}

@article{inubushi2017reservoir,
  title={Reservoir computing beyond memory-nonlinearity trade-off},
  author={Inubushi, Masanobu and Yoshimura, Kazuyuki},
  journal={Scientific reports},
  volume={7},
  number={1},
  pages={10199},
  year={2017},
  publisher={Nature Publishing Group UK London}
}

@article{hauser2011towards,
  title={Towards a theoretical foundation for morphological computation with compliant bodies},
  author={Hauser, Helmut and Ijspeert, Auke J and F{\"u}chslin, Rudolf M and Pfeifer, Rolf and Maass, Wolfgang},
  journal={Biological cybernetics},
  volume={105},
  number={5},
  pages={355--370},
  year={2011},
  publisher={Springer}
}

@article{nakajima2013soft,
  title={A soft body as a reservoir: case studies in a dynamic model of octopus-inspired soft robotic arm},
  author={Nakajima, Kohei and Hauser, Helmut and Kang, Rongjie and Guglielmino, Emanuele and Caldwell, Darwin G and Pfeifer, Rolf},
  journal={Frontiers in computational neuroscience},
  volume={7},
  pages={91},
  year={2013},
  publisher={Frontiers Media SA}
}

@article{urbain2017morphological,
  title={Morphological properties of mass--spring networks for optimal locomotion learning},
  author={Urbain, Gabriel and Degrave, Jonas and Carette, Benonie and Dambre, Joni and Wyffels, Francis},
  journal={Frontiers in neurorobotics},
  volume={11},
  pages={16},
  year={2017},
  publisher={Frontiers Media SA}
}

@article{barazani2020microfabricated,
  title={Microfabricated neuroaccelerometer: integrating sensing and reservoir computing in MEMS},
  author={Barazani, Bruno and Dion, Guillaume and Morissette, Jean-Fran{\c{c}}ois and Beaudoin, Louis and Sylvestre, Julien},
  journal={Journal of Microelectromechanical Systems},
  volume={29},
  number={3},
  pages={338--347},
  year={2020},
  publisher={IEEE}
}

@article{sun2021novel,
  title={Novel nondelay-based reservoir computing with a single micromechanical nonlinear resonator for high-efficiency information processing},
  author={Sun, Jie and Yang, Wuhao and Zheng, Tianyi and Xiong, Xingyin and Liu, Yunfei and Wang, Zheng and Li, Zhitian and Zou, Xudong},
  journal={Microsystems \& Nanoengineering},
  volume={7},
  number={1},
  pages={83},
  year={2021},
  publisher={Nature Publishing Group UK London}
}

@inproceedings{zhang2022harnessing,
  title={Harnessing physical reservoir computing in nonlinear mechanical metastructures},
  author={Zhang, Yuning and Wang, Kon-Well},
  booktitle={AIAA Scitech 2022 Forum},
  pages={1464},
  year={2022}
}

@article{shougat2024multiplex,
  title={Multiplex-free physical reservoir computing with an adaptive oscillator},
  author={Shougat, Md Raf E Ul and Li, XiaoFu and Perkins, Edmon},
  journal={Physical Review E},
  volume={109},
  number={2},
  pages={024203},
  year={2024},
  publisher={APS}
}

@article{soures2019deep,
  title={Deep liquid state machines with neural plasticity for video activity recognition},
  author={Soures, Nicholas and Kudithipudi, Dhireesha},
  journal={Frontiers in neuroscience},
  volume={13},
  pages={686},
  year={2019},
  publisher={Frontiers Media SA}
}

@article{pixabay_fish,
  author        = {Rethinktwice},
  title         = {Fish, Carassius, Veiltail},
  howpublished  = {\url{https://pixabay.com/photos/fish-carassius-veiltail-fins-5917864/}},
  year          = {2021},
  note          = {Pixabay, accessed June 2026}
}
\bibliographystyle{spiebib} 

\end{document}